\documentclass[aps, prd, twocolumn, 
showpacs, showkeys, 
print, preprintnumbers, nofootinbib,
superscriptaddress 
]{revtex4-2}

\usepackage[utf8]{inputenc}
\usepackage[T1]{fontenc}
\usepackage{lmodern}
\usepackage{microtype}
\usepackage[english]{babel}

\usepackage{amssymb,amsbsy,amsmath,amsfonts,amsthm}
\allowdisplaybreaks
\usepackage{yhmath} 

\usepackage{slashed}
\usepackage{bm}
\usepackage{times}
\usepackage{soul}
\setstcolor{red}

\usepackage[dvipsnames,table,xcdraw]{xcolor}

\usepackage{graphicx}
\usepackage{epstopdf}
\usepackage{tabularx}
\usepackage{multirow}
\usepackage{longtable}

\usepackage{hyperref}
\hypersetup{ colorlinks=true, linkcolor=blue, citecolor=cyan, urlcolor=red
}

\usepackage{url}
\usepackage{orcidlink}

\usepackage{comment}

\begin{document}

\title{Phenomenological model for $\gamma\gamma^* \to K\bar{K}^*(892)$ constraining the $f_1(1420)$ transition form factor}

\author{Xiu-Lei Ren\orcidlink{0000-0002-5138-7415}}
\affiliation{Helmholtz Institut Mainz,   D-55099 Mainz, Germany }

\author{Igor Danilkin\orcidlink{0000-0001-8950-0770}}
\affiliation{Institut f\"ur Kernphysik \& PRISMA$^+$  Cluster of Excellence, Johannes Gutenberg Universit\"at,  D-55099 Mainz, Germany}

\author{Marc Vanderhaeghen\orcidlink{0000-0003-2363-5124}}
\affiliation{Institut f\"ur Kernphysik \& PRISMA$^+$  Cluster of Excellence, Johannes Gutenberg Universit\"at,  D-55099 Mainz, Germany}

\begin{abstract}
We present a phenomenological study of the $\gamma\gamma^*\to K\bar{K}^*(892)$ process by including the $s$-channel production of the $\eta(1475)$ and $f_1(1420)$ resonances. The non-resonant channel via $K$- and $K^*$-exchanges is investigated carefully by performing the Lorentz tensor decomposition and is constructed to yield a correct high-energy Regge behavior. The transition form factor of $f_1(1420)$ is adjusted to achieve a reasonable description of the existing L3 data in the $f_1(1420)$ resonance region. 
This model is intended to serve as a Monte Carlo generator for the analysis currently being performed by the BESIII Collaboration. We also estimate the polarized  $\gamma\gamma^*\to K\bar{K}^*(892)$ cross sections and demonstrate how to extract the transition form factors of $f_1(1420)$ from the polarized cross sections.
\end{abstract}


\maketitle

\date{\today}


\section{Introduction}
The electromagnetic transition form factors (TFFs) of mesons, accessed through the fusion of two (virtual) photons into a meson, comprise the inner-structural information of the hadrons. Studying the low momentum behavior of these TFFs can deepen our understanding of the quark structure of these mesons in nonperturbative QCD.  In the case of axial-vector mesons ($A$), the $\gamma\gamma \to A$ process is forbidden due to the Landau-Yang theorem~\cite{Landau:1948kw,Yang:1950rg}. However, the measurement of their TFFs can be accessed through singly-virtual or doubly-virtual processes. Currently, several measurements focus on the space-like process $e^+e^-\to e^+e^- A$ with $A=f_1(1285)$ and $f_1(1420)$~\cite{Gidal:1987bn,Gidal:1987bm,TPCTwoGamma:1988izb,TPC-TWO-GAMMA:1988aph,L3:2001cyf,L3:2007obw} via the singly-virtual $\gamma\gamma^*\to A$ production. The precision of the existing data makes it challenging to accurately determine the TFFs of $f_1(1285)$ and $f_1(1420)$, particularly regarding their momentum dependence. Phenomenologically, a commonly used parameterization of $f_1$ TFFs is the dipole form~\cite{L3:2001cyf,L3:2007obw,Pauk:2014rta}, which relies on the quark model~\cite{Schuler:1997yw}. The $f_1$ TFFs constrained by the large-$N_c$ and operator product expansion arguments are proposed by Melnikov and Vainshtein~\cite{Melnikov:2003xd}, and are further anti-symmetrized to satisfy the Landau-Yang suppression in Ref.~\cite{Jegerlehner:2017gek}. Most recently, the $f_1(1285)$ and $f_1(1420)$ TFFs have been studied using the resonance chiral theory~\cite{Roig:2019reh} and the holographic models~\cite{Leutgeb:2019gbz}. When the virtuality becomes very large, the asymptotic behavior of $f_1$ TFFs is obtained from the light-cone expansion~\cite{Hoferichter:2020lap}. Incorporating this constraint, a vector meson dominance inspired parameterization of the $f_1(1285)$ TFFs have been proposed in Refs.~\cite{Zanke:2021wiq,Hoferichter:2023tgp} through a global fit of all relevant data. However, the determination of the $f_1(1420)$ TFFs remains inconclusive due to the limited experimental data. A better determination of the $f_1(1285)$ and $f_1(1420)$ TFFs is also timely in view of its contribution to the hadronic light-by-light (HLbL) contribution to the muon's $g-2$~\cite{Aoyama:2020ynm,Aoyama:2012wk,Aoyama:2019ryr,Czarnecki:2002nt,Gnendiger:2013pva,Davier:2017zfy,Keshavarzi:2018mgv,Colangelo:2018mtw,Hoferichter:2019mqg,Davier:2019can,Keshavarzi:2019abf,Kurz:2014wya,Melnikov:2003xd,Masjuan:2017tvw,Colangelo:2017fiz,Hoferichter:2018kwz,Gerardin:2019vio,Bijnens:2019ghy,Colangelo:2019uex,Blum:2019ugy,Colangelo:2014qya}.

To improve upon the experimental situation, the BESIII Collaboration conducted a preliminary analysis of the $\gamma\gamma^*\to K^\pm K^{*\mp}(892) \to K^+K^-\pi^0$ reaction~\cite{Effenberger_MasterThesis}. The $\gamma\gamma^*\to K\bar{K}^*(892)$ process is considered as an ideal channel to extract the $f_1(1420)$ TFFs, since the $K\bar{K}^*+\mathrm{c.c.}$ channel is the dominant decay mode of the $f_1(1420)$ state with a branching ratio around $96\%$~\cite{WA102:1998zhh,ParticleDataGroup:2022pth}. In the BESIII preliminary data analysis, the ``GaGaRes'' Monte Carlo (MC) program~\cite{Berends:2001ta} was used to distinguish between different resonance contributions, i.e. $\eta(1475)$ and $f_1(1420)$, to the $\gamma\gamma^*\to K\bar{K}^*(892)$ process. However, the GaGaRes program does not implement interference between these amplitudes, nor does it include non-resonant mechanisms. These limitations could potentially lead to misinterpretations of the data.

In order to provide a more realistic MC generator for the data analysis of the BESIII measurement, we propose a phenomenological model for  $\gamma\gamma^*\to K\bar{K}^*(892)$ with the charged $K$ and $K^*$ final states. This model includes $s$-channel contributions from the $\eta(1475)$ and $f_1(1420)$ production mechanisms. Additionally, we consider non-resonant contributions via the $K$- and $K^*$-exchange channels. The corresponding amplitude is constructed to exhibit the correct high-energy behavior using the Reggeized exchanges of $K$ and $K^*$ mesons. To determine the interference effects among the different contributions, we utilize available L3 experimental data on $e^+e^- \to e^+e^- K_S^0 K^\pm \pi^\mp$~\cite{L3:2007obw}.
Subsequently, we predict polarized (differential) cross sections of $\gamma\gamma^*\to K\bar{K}^*$ within the $Q^2$ regime of the BESIII experiment.

The paper is organized as follows: In Sec.~\ref{SecII}, we present the $\gamma\gamma^* \to K\bar{K}^*(892)$ amplitude in our phenomenological model. To constrain the $f_1(1420)$ TFFs, in Sec.~\ref{SecIII}, we describe the existing L3 data and present the prediction of the polarized (differential) cross section of $\gamma\gamma^* \to K\bar{K}^*(892)$ in the BESIII energy range. The extraction of TFFs is also discussed there. Finally, we summarize the main results in Sec.~\ref{SecIV}. Some technical details are given in two appendices.

\begin{figure*}[t]
\includegraphics[width=0.66\textwidth]{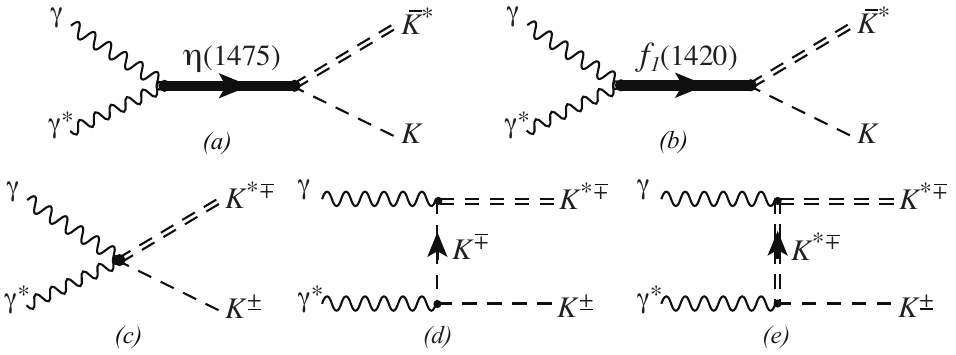}
  \caption{Tree level diagrams of the $\gamma\gamma^* \to K\bar{K}^*(892)$ reaction via the $s$-channel $\eta(1475)$ and $f_1(1420)$ production, the contact term, and the $t$-channel charged $K$ and $\bar{K}^*$ exchange. The diagrams with crossed photon lines are not shown but are included in the calculation.}
  \label{Fig:diags}
\end{figure*}

\section{Theoretical framework}\label{SecII}
To parametrize the $\gamma\gamma^* \to K\bar{K}^*(892)$ amplitude, one needs to account for the contributions of $\eta(1475)$ and $f_1(1420)$ resonances, as shown in Fig.~\ref{Fig:diags}-(a,b). The $s$-channel production of $\eta(1475)$ via two-photon fusion is allowed, and the $K\bar{K}^*$ channel is the major decay mode of the $\eta(1475)$ state. For the real photon fusion process, $\eta(1475)$ is expected to be the dominant resonant process, since the production of an axial-vector resonance by real photons is forbidden by the Landau-Yang theorem~\cite{Landau:1948kw,Yang:1950rg}. If one photon is virtual, the axial-vector mesons are allowed to be produced in the photon-photon fusion. Thus, we have an $s$-channel contribution to the $\gamma\gamma^*\to K\bar{K}^*(892)$ reaction via the production of the $f_1(1420)$ resonance. Besides, in Fig.~\ref{Fig:diags}, the non-resonant channels, namely $t$- and $u$-channel $K,~K^*$ exchanges mechanisms, along with the associated contact terms to ensure the electromagnetic gauge invariance, are included at the tree level. In our model, the total amplitude of the $\gamma\gamma^*\to K\bar{K}^*$ reaction is written as 
\begin{equation}
   \mathcal{M}_{\gamma\gamma^*\to K\bar{K}^*} = \mathcal{M}_{\bar{\eta}} + \mathcal{M}_{f_1} + \mathcal{M}_\text{non-res.},
\end{equation}
where we use shorthand notation $\bar{\eta}$ to denote the $\eta(1475)$ state in the following.

\subsection{$\gamma\gamma^*\to \eta(1475)\to K\bar{K}^*(892)$ channel}
The production of $\eta(1475)$ resonance 
by two photons, $\gamma^*(q_1,\lambda_1)+\gamma^*(q_2,\lambda_2)\to \eta(1475)$, is described by the matrix element, 
\begin{equation}
\begin{aligned}
  \mathcal{M}_{\bar{\eta}\gamma^*\gamma^*}(\lambda_1,\lambda_2)  = &-i e^2 \epsilon_{\mu\nu\alpha\beta}\,\varepsilon^{\mu}(q_1,\lambda_1)\,\varepsilon^{\nu}(q_2,\lambda_2) \\
  & \times q_1^\alpha \, q_2^\beta \, F_{\bar{\eta}\gamma^*\gamma^*}(Q_1^2, Q_2^2),
\end{aligned}
\end{equation}
where the polarization vectors of (real and virtual) photons are denoted as $\varepsilon^\mu(q_i,\lambda_i)$ with $\lambda_{1,2}=0,\,\pm 1$. The structure information of the $\eta(1475)$ state is encoded in the  space-like $\gamma^*\gamma^*$ TFF, which is taken in the monopole form 
\begin{equation}
  F_{\bar{\eta}\gamma^*\gamma^*}(0, Q_2^2) = \frac{F_{\bar{\eta}\gamma^*\gamma^*}(0,0)}{1+Q_2^2/\Lambda_{\bar{\eta}}^2},	
\end{equation}
with $\Lambda_{\bar{\eta}}=1470$ MeV from Ref.~\cite{L3:2007obw}. 
The TFF at $Q_1^2=Q_2^2=0$, $F_{\bar{\eta}\gamma^*\gamma^*}(0,0)$, is related to the decay width of $\eta(1475)\to \gamma\gamma$, 
\begin{equation}
\Gamma_{\bar{\eta}\to\gamma\gamma} = \frac{\pi\alpha^2}{4} M_{\bar{\eta}}^3 |F_{\bar{\eta}\gamma^*\gamma^*}(0,0)|^2, 	
\end{equation}
with the $\eta(1475)$ mass $M_{\bar{\eta}}=1475$ MeV and the fine-structure constant $\alpha=e^2/(4\pi)\approx 1/137$. 
Since the decay modes of the $\eta(1475)$ state are not well established in the experiment, we assume that the total width of $\eta(1475)$, $\Gamma_{\bar{\eta}}=90\pm 9$ MeV, is obtained by the sum of the $\eta(1475)\to K\bar{K}^*+\mathrm{c.c.}$, $\eta(1475)\to a_0(980)\pi^0$, and $\eta(1475)\to \gamma\gamma$ channels, 
as listed in PDG~\cite{ParticleDataGroup:2022pth}. Using the determined branching ratio by the L3 Collaboration~\cite{L3:2007obw}, we obtain the decay width of $\eta(1475)\to\gamma\gamma$ as $\Gamma_{\bar{\eta}\to\gamma\gamma}\simeq 0.23~\mathrm{keV}$. This leads to the value of the normalization of  the TFF $F_{\bar{\eta}\gamma^*\gamma^*}(0,0)\simeq 0.0414$ GeV$^{-1}$.

The effective Lagrangian to describe the interaction of $\eta(1475) K^\pm K^{*\mp}$ is written as 
\begin{equation}
\begin{aligned}
  \mathcal{L}_{\bar{\eta} KK^*} =& i g_{\bar{\eta} KK^*}\Bigl[\bar{\eta} \bigl( \partial_\mu K^- K^{*+,\mu} - \partial_\mu K^+ K^{*-,\mu}\bigr) \\
  & - \partial_\mu\bar{\eta}\bigl(  K^- K^{*+,\mu}  - K^+ K^{*-,\mu}\bigr)\Bigr],	
\end{aligned}
\end{equation}
where the charged $K^{(*)\pm}$ fields denote the annihilation of $K^{(*)\mp}$ and creation of $K^{(*)\pm}$ particles.
The dimensionless coupling $g_{\bar{\eta} KK^*}$ is fixed by the partial decay width of $\eta(1475)\to K\bar{K}^*$, 
\begin{equation}
  \Gamma_{\bar{\eta}\to K\bar{K}^*} =\frac{1}{2\pi} g_{\bar{\eta} KK^*}^2 \frac{\left[q_{\bar{\eta}\to K\bar{K}^*}(M_{\bar{\eta}}^2)\right]^3}{M_{K^*}^2},	
\end{equation}
with the momentum in the rest frame of $\eta(1475)\to K\bar{K}^*$ channel, 
\begin{equation}  
q_{\bar{\eta}\to K\bar{K}^*}(W^2)=\frac{\lambda^{1/2}(W^2, M_{K^*}^2, m_K^2)} {2W} ,
\end{equation}	
where $\lambda$ denotes the K\"{a}ll\'{e}n triangle function, $\lambda(x,y,z)\equiv x^2+y^2+z^2-2xy-2xz-2yz$, and $W$ is the total energy of the $K\bar{K}^*$ system. To estimate the magnitude of $\Gamma_{\bar{\eta}\to K\bar{K}^*}$, we assume $$\Gamma_{\bar{\eta}\to K\bar{K}\pi} = \Gamma_{\bar{\eta}\to K\bar{K}^* + \mathrm{c.c.}}+ \Gamma_{\bar{\eta}\to a_0\pi^0}\approx \Gamma_{\bar{\eta}} = 90~\mathrm{MeV}.$$  
Using the available measurements of the branching ratio, $\Gamma_{\bar{\eta}\to K\bar{K}^* + \mathrm{c.c.}}/\Gamma_{\bar{\eta}\to K\bar{K}\pi} = 0.5$~\cite{Baillon:1967zz} or~$<0.25$~\cite{Edwards:1982nc}, we obtain the maximum value of $\Gamma_{\bar{\eta}\to K\bar{K}^*+\mathrm{c.c.}}\simeq 45$ MeV, which results in the value of $g_{\bar{\eta}KK^*}\simeq 2.04$.

Finally, the tree-level amplitude of $\gamma\gamma^*\to \eta(1475) \to K\bar{K}^*(892)$ reaction is 
\begin{equation}\label{Eq:AmpEta}
\begin{aligned}
  \mathcal{M}_{\bar{\eta}} & = -2\,i\,e^2\,g_{\bar{\eta} KK^*} F_{\bar{\eta}}(0,Q_2^2)\,\epsilon_{\mu\nu\sigma\beta} \\
  &\quad \times \varepsilon^{\mu}(q_1,\lambda_1)\,\varepsilon^\nu(q_2,\lambda_2)\varepsilon^{*\alpha}(p_1,\Lambda_{K^*})\\
  &\quad \times  \frac{(q_1)^\sigma (q_2)^\beta {(p_2)}_\alpha}{(q_1+q_2)^2 - M_{\bar{\eta}}^2 + i M_{\bar{\eta}} \Gamma_{\bar{\eta}}(W^2)} \\
  &\quad \times \left(\frac{D_1[q_{\bar{\eta}\to\gamma\gamma^*}(W^2)\,R_{\bar{\eta}}]}{D_1[q_{\bar{\eta}\to\gamma\gamma^*}(M_{\bar{\eta}}^2)\,R_{\bar{\eta}}]}\right)^{1/2}\\
  &\quad \times 
  \left(\frac{D_1[q_{\bar{\eta}\to K\bar{K}^*}(W^2)\,R_{\bar{\eta}}]}{D_1[q_{\bar{\eta}\to K\bar{K^*}}(M_{\bar{\eta}}^2)\,R_{\bar{\eta}}]}\right)^{1/2},	
\end{aligned}
\end{equation}
where the energy-dependent width of $\eta(1475)$ is written as 
\begin{equation}
\begin{aligned}
  \Gamma_{\bar{\eta}}(W^2) &= \Gamma_{\bar{\eta}}(M_{\bar{\eta}}^2) \Biggl\{ \mathrm{Br}(\bar{\eta}\to KK^*) \frac{M_{\bar{\eta}}}{W} \\
  & \times \left[\frac{q_{\bar{\eta}\to K\bar{K}^*}(W^2)}{q_{\bar{\eta}\to K\bar{K}^*}(M_{\bar{\eta}}^2)} \right]^3  \frac{D_1[q_{\bar{\eta}\to K\bar{K}^*}(W^2)\, R_{\bar{\eta}}]}{D_1[q_{\bar{\eta}\to K\bar{K}^*}(M_{\bar{\eta}}^2)\,R_{\bar{\eta}}]} \\
  &\times \Theta\left(W^2-(m_K+M_{K^*})^2\right) \\
  & + \mathrm{Br}(\bar{\eta}\to a_0\pi^0) \frac{M_{\bar{\eta}}}{W} \frac{q_{\bar{\eta}\to a_0\pi^0}(W^2)}{q_{\bar{\eta}\to a_0\pi^0}(M_{\bar{\eta}}^2)} \\
  & \times \Theta\left(W^2-(m_\pi+M_{a_0})^2\right) \Biggr\},
\end{aligned}
\end{equation}
with the Blatt-Weisskopf barrier factor~\cite{Blatt_Weisskopf_book}~$D_1(x)=1/(1+x^2)$, and the momentum 
\begin{equation} 
	q_{\bar{\eta}\to \gamma\gamma^*}(W^2)=\frac{\lambda^{1/2}(W^2,0, Q_2^2)}{2W}.
\end{equation}
The barrier effective radius $R_{\bar{\eta}}$ for the $\eta(1475)$ resonance, accounting for finite size effects, is usually taken from $1$ to $7$ GeV$^{-1}$~\cite{Belle:2009xpa}. Since the results are found to display very little sensitivity to this value in the kinematical region studied here, then we set $R_{\bar{\eta}} =3.0$ GeV$^{-1}$ as used in Ref.~\cite{BESIII:2021dot}.

\subsection{$\gamma\gamma^*\to f_1(1420) \to K\bar{K}^*(892)$ channel}
The production of the $f_1(1420)$ resonance by two-photon fusion is allowed when one or both photons are virtual. The amplitude of $\gamma^*(q_1,\lambda_1)+\gamma^*(q_2,\lambda_2)\to f_1(1420)$ can be parametrized by three structures~\cite{Poppe:1986dq,Schuler:1997yw},
\begin{align}\label{Eq:Amp_gagaf1}
\mathcal{M}_{f_1\gamma^*\gamma^*}
& =i e^2\varepsilon_\mu(q_1,\lambda_1)\varepsilon_\nu(q_2,\lambda_2)\varepsilon^{\omega*}(q_1+q_2,\Lambda_{f_1}) \nonumber\\
&\times \epsilon_{\rho\sigma\tau\omega} \biggl[R^{\mu\rho}(q_1,q_2) R^{\nu\sigma}(q_1,q_2) (q_1-q_2)^\tau \nonumber\\
&\qquad \times \frac{\nu}{M_{f_1}^2} F^{TT}_{f_1\gamma^*\gamma^*}(Q_1^2,Q_2^2) \nonumber\\
	&\quad + R^{\nu\rho}(q_1,q_2)\biggl( q_1^\mu + \frac{Q_1^2}{\nu}q_2^\mu\biggr) q_1^\sigma q_2^\tau \\
 &\qquad \times \frac{1}{M_{f_1}^2} F^{LT}_{f_1\gamma^*\gamma^*}(Q_1^2,Q_2^2) \nonumber\\
	&\quad + R^{\mu\rho}(q_1,q_2) \biggl(q_2^\nu + \frac{Q_2^2}{\nu}q_1^\nu\biggr) q_2^\sigma q_1^\tau \nonumber\\
 &\qquad \times \frac{1}{M_{f_1}^2} F^{TL}_{f_1\gamma^*\gamma^*}(Q_1^2, Q_2^2) \biggr]\nonumber,
\end{align}	
where the symmetric transverse tensor is defined as 
\begin{equation}
\begin{aligned}
  R^{\mu\nu}(q_1,q_2) &= -g^{\mu\nu} + \frac{1}{X} \Biggl[ \nu(q_1^\mu q_2^\nu + q_2^\mu q_1^\nu) \\
  &\qquad + Q_1^2 q_2^\mu q_2^\nu + Q_2^2 q_1^\mu q_1^\nu \Biggr],
\end{aligned}
\end{equation}
with the virtual photon flux factor, $X=(q_1\cdot q_2)^2 - q_1^2 q_2^2 = \nu^2 -Q_1^2Q_2^2$, and $\nu = q_1\cdot q_2$. 
The structure information of the $f_1(1420)$ state is encoded in the  three TFFs $F_{f_1\gamma^*\gamma^*}^{TT,TL,LT}$, which are functions of the virtualities of both photons. The superscript $TT$ indicates the two transverse photons, while $TL$ ($LT$) stands for the first photon being transverse (longitudinal) and the second photon being longitudinal (transverse). The TFFs $F^{LT}$ and $F^{TL}$ are related as 
\begin{equation}
	F^{LT}_{f_1\gamma^*\gamma^*}(Q_1^2,Q_2^2) = F^{TL}_{f_1\gamma^*\gamma^*}(Q_2^2,Q_1^2).
\end{equation}

In the current work, we consider the process with one real photon and take $Q_1^2=0$. The amplitude of $\gamma(q_1,\lambda_1)+\gamma^*(q_2,\lambda_2)\to f_1(1420)$ is then expressed as~\cite{Pascalutsa:2012pr}
\begin{equation}\label{Eq:Amp_gagaf1_Q2}
\begin{aligned}
& \mathcal{M}_{f_1\gamma\gamma^*} 
 =i e^2\varepsilon_\mu(q_1,\lambda_1)\varepsilon_\nu(q_2,\lambda_2)\varepsilon^{\omega*}(q_1+q_2,\Lambda_{f_1}) \\
&\times \epsilon_{\rho\sigma\tau\omega} \Biggl\{ \biggl[ \nu g^{\mu\rho} g^{\nu\sigma} (q_1-q_2)^\tau 
-  g^{\nu\rho} q_2^\mu  q_1^\sigma q_2^\tau \\
&+ g^{\mu\rho}  \left(q_2^\nu  + q_1^\nu + \frac{Q_2^2}{\nu} q_1^\nu \right) q_1^\sigma q_2^\tau \biggr]  \frac{1}{M_{f_1}^2} F^{TT}_{f_1\gamma^*\gamma^*}(0,Q_2^2) \\
	&+  g^{\mu\rho} \biggl(q_2^\nu +\frac{Q_2^2}{\nu} q_1^\nu \biggr) q_1^\sigma q_2^\tau  \frac{1}{M_{f_1}^2} F^{TL}_{f_1\gamma^*\gamma^*}(0, Q_2^2) \Biggr\},
\end{aligned}
\end{equation}
which results in the helicity amplitudes of $\gamma\gamma^*\to f_1(1420)$ 
\begin{equation}\label{Eq:gagaF1Amp}
\begin{aligned}
	\mathcal{M}_{f_1\gamma\gamma^*}^{\lambda_1=\lambda_2=\pm1, \Lambda_{f_1}=0} & = -e^2 \frac{\nu\, Q_2^2}{M_{f_1}^3}  F_{f_1\gamma^*\gamma^*}^{TT}(0,Q_2^2), \\
	\mathcal{M}_{f_1\gamma\gamma^*}^{\lambda_1=-1, \lambda_2=0, \Lambda_{f_1}=0}  & = -e^2\frac{\nu\,Q_2}{M_{f_1}^2} F_{f_1\gamma^*\gamma^*}^{TL}(0,Q_2^2).	
\end{aligned}
\end{equation}
Note that the third TFF, $F_{f_1\gamma^*\gamma^*}^{LT}(0,Q_2^2)$, decouples in the single virtual case~\footnote{$F_{f_1\gamma^*\gamma^*}^{LT}(0,Q_2^2)$ can be extracted from the interference observables $\tau_{TL}$ or $\tau_{TL}^a$, as discussed in Ref.~\cite{Pascalutsa:2012pr}.}.  The remaining two TFFs can be independently determined from the polarized cross sections, $\sigma_{TT}$ and $\sigma_{TL}$, of the $\gamma\gamma^*\to f_1(1420)$ resonance production process,
\begin{equation}\label{Eq:XSec_TFFs}
\begin{aligned}
	\sigma_{TT}(Q_2^2) &=  \frac{2\pi^2 \alpha^2}{M_{f_1}\Gamma_{f_1}} \frac{Q_2^4}{M_{f_1}^4} \left(1+\frac{Q_2^2}{M_{f_1}^2}\right) \\
	&\quad \times \bigl[ F^{TT}_{f_1\gamma^*\gamma^*} (0, Q_2^2)\bigr]^2,\\
	\sigma_{TL}(Q_2^2) &= \frac{4\pi^2 \alpha^2}{M_{f_1}\Gamma_{f_1}} \frac{Q_2^2}{M_{f_1}^2}\left(1+\frac{Q_2^2}{M_{f_1}^2}\right) \\
	&\quad \times \bigl[ F^{TL}_{f_1\gamma^*\gamma^*} (0,Q_2^2)\bigr]^2.
\end{aligned}
\end{equation}
Note that $\sigma_{TT}$ is suppressed by $Q_2^2/(2M_{f_1}^2)$ in comparison with $\sigma_{TL}$, in the low $Q_2^2$ region, for the case where the TFF  $F^{TT}_{f_1\gamma^*\gamma^*} (0,Q_2^2)$ and  $F^{TL}_{f_1\gamma^*\gamma^*} (0,Q_2^2)$ are of similar magnitude, as discussed further on.

The relations of Eq.~\eqref{Eq:XSec_TFFs} are strictly valid for the resonance production process at the resonance position. The model for the process $\gamma\gamma^*\to K\bar{K}^*(892)$ developed in this work will allow to quantify the resonance dominance, and enable to extract the TFF by a fit to the $\gamma\gamma^*\to K\bar{K}^*(892)$ data at the resonance position. 
For this purpose, we need to work out the matrix element of $f_1(1420)$ decay to $K\bar{K}^*$. The corresponding 
effective Lagrangian of $f_1(1420)K^+ K^{*-}$ vertex is given by 
\begin{equation}
\begin{aligned}
   \mathcal{L}_{f_1KK^{*}} &= \frac{g_{f_1 KK^{*}}}{M_{f_1}} (\partial_\mu {K^{*+}}_\nu - \partial_\nu {K^{*+}}_\mu)  \\
   &\quad \times \bigl(\partial^\mu (f_1)^\nu - \partial^\nu (f_1)^\mu\bigr) \, K^-,	
\end{aligned}
\end{equation}
where the coupling $g_{f_1KK^*}$ is determined by the decay width of $f_1(1420)\to K^+K^{*-}$,
\begin{equation}
\begin{aligned}
  \Gamma_{f_1\to K^+K^{*-}}	& = \frac{g_{f_1 KK^*}^2 }{12\pi M_{f_1}^4} q_{f_1\to KK^{*}}(M_{f_1}^2) \\
  &\times \biggl[ 2M_{K^*}^2 M_{f_1}^2 + (M_{f_1}^2+M_{K^*}^2-m_K^2)^2\biggr],
\end{aligned}
\end{equation}
where the rest frame momentum of $f_1\to K\bar{K}^*$ channel $q_{f_1\to K\bar{K}^*}(W^2)$ has the same functional form as $q_{\bar{\eta}\to K\bar{K}^*}(W^2)$.	
According to the branching ratio of 
$$\mathrm{Br}(f_1(1420)\to K\bar{K}^* + \mathrm{c.c.}) = 96.0\pm1.0\pm1.0\%$$ measured by WA102 Collaboration~\cite{WA102:1998zhh}, we have $\Gamma_{f_1\to K\bar{K}^* + \mathrm{c.c.}}=52.32$ MeV, which leads to the coupling value $g_{f_1 KK^*}=1.027$. 

Using the above vertices, the $s$-channel amplitude of $\gamma(q_1)\gamma^*(q_2)\to \bar{K}^*(p_1) K(p_2)$ via the $f_1(1420)$ state [Fig.~\ref{Fig:diags}-(b)] can be written as 
\begin{equation}\label{Eq:Ampf1}
\begin{aligned}
	\mathcal{M}_{f_1} &=  \frac{2\,i\,e^2\, g_{f_1 KK^*}}{M_{f_1}} \varepsilon_{\mu}(q_1,\lambda_1) \varepsilon_{\nu}(q_2,\lambda_2) \varepsilon_{\alpha}^*(p_1,\Lambda_{K^*}) \\
	&\times \frac{(p_1\cdot p_{f_1}) g^{\alpha \beta} - (p_1)^\beta (p_{f_1})^\alpha}{p_{f_1}^2-M_{f_1}^2 + i M_{f_1}\Gamma_{f_1}(W^2)} \\
	&\times \epsilon_{\rho\sigma\tau\beta}  \Biggl\{ \biggl[ \nu g^{\mu\rho} g^{\nu\sigma} (q_1-q_2)^\tau 
-  g^{\nu\rho} q_2^\mu  q_1^\sigma q_2^\tau \\
&\quad + g^{\mu\rho}  (q_2^\nu + q_1^\nu  + \frac{Q_2^2}{\nu} q_1^\nu) q_1^\sigma q_2^\tau \biggr] \\
&\quad \times \frac{1}{M_{f_1}^2} F^{TT}_{f_1\gamma^*\gamma^*}(0,Q_2^2) \\
	&+  g^{\mu\rho} \biggl(q_2^\nu +\frac{Q_2^2}{\nu} q_1^\nu \biggr) q_1^\sigma q_2^\tau  \frac{1}{M_{f_1}^2} F^{TL}_{f_1\gamma^*\gamma^*}(0, Q_2^2) \Biggr\},
\end{aligned}	
\end{equation}
where the $f_1(1420)$ momentum  is $p_{f_1}=q_1+q_2$. The energy-dependent width of the intermediate $f_1(1420)$ resonance is introduced in the propagator  
\begin{equation}
\begin{aligned}
  \Gamma_{f_1}(W^2) &= \Gamma_{f_1}(M_{f_1}^2) \Biggl[ \mathrm{Br}(f_1\to KK^*) \frac{M_{f_1}}{W} \frac{q_{f_1\to KK^*}(W^2)}{q_{f_1\to KK^*}(M_{f_1}^2)} \\
  &\quad \times \Theta(W^2-(M_{K^*}+m_K)^2)  \Biggr].
\end{aligned}
\end{equation}

In order to extract the TFFs $F_{f_1\gamma^*\gamma^*}^{TT}(0,Q_2^2)$ and $F_{f_1\gamma^*\gamma^*} ^{TL}(0,Q_2^2)$ in Eq.~\eqref{Eq:Ampf1}, one needs the polarized cross section for the $\gamma\gamma^*\to K\bar{K}^*(892)$ reaction. Currently only the helicity averaged data is available, as measured by the L3 Collaboration~\cite{L3:2007obw}. To estimate the smaller $\sigma_{TT}$ contribution to the cross-section, which is suppressed as $Q_2^2/(2M_{f_1}^2)$ compared to $\sigma_{TL}$ at low $Q_2^2$, we will use the quark model estimate~\cite{Cahn:1986qg}, which relates both TFFs as: 
\begin{equation}\label{Eq:TFFquarkmodel}
   F_{f_1\gamma^*\gamma^*}^{TT}(0,Q_2^2) = - F_{f_1\gamma^*\gamma^*}^{TL}(0,Q_2^2) .
\end{equation} 
Therefore, the matrix element~(Eq.~\eqref{Eq:Amp_gagaf1_Q2}) of the $\gamma(q_1,\lambda_1)\gamma^*(q_2,\lambda_2)\to f_1(1420)$ reaction  depends solely on one dominant form factor, $F_{f_1\gamma^*\gamma^*}^{TL}(0,Q_2^2)$, which is parameterized by the dipole form, 
\begin{equation}\label{Eq:TFFf1}
  F_{f_1\gamma^*\gamma^*}^{TL}(0, Q_2^2) = \frac{F_{f_1\gamma^*\gamma^*}^{TL}(0,0)}{\bigl(1+Q_2^2/\Lambda_{f_1}^2\bigr)^2}\,. 
\end{equation}
This parameterization shares the same functional form as the one proposed
in the non-relativistic quark model~\cite{Schuler:1997yw} with $\Lambda_{f_1}=M_{f_1}$. The dipole form has been demonstrated to satisfy the large $Q_2$ asymptotic $1/Q_2^4$ behavior in Ref.~\cite{Hoferichter:2020lap}. However, there is no compelling reason to identify $\Lambda_{f_1}$ with $M_{f_1}$. Instead, this value can be adjusted to better describe the L3 data in the energy range of $f_1(1420)$.
Regarding the normalization of the TFF, $F_{f_1\gamma^*\gamma^*}^{TL}(0,0)$, it is conventional to define an equivalent two-photon decay width of $f_1(1420)$ as~\cite{TPCTwoGamma:1988izb}
\begin{equation}
\begin{aligned}
 \tilde{\Gamma}_{f_1\to\gamma\gamma} & \equiv \lim_{Q_2^2\to 0}  \frac{M_{f_1}^2}{2Q_2^2} \Gamma(f_1\to \gamma_T \gamma_L^*) \\
& = \frac{\pi\alpha^2}{4}M_{f_1} \frac{1}{3} [F_{f_1\gamma^*\gamma^*}^{TL}(0,0)]^2,
\end{aligned}
\end{equation}
via the decay width $\Gamma(f_1(1420)\to\gamma_T \gamma^*_L)$ into  a real transverse photon ($Q_1^2=0$) and a quasi-real longitudinal photon with virtuality ($Q_2^2\to 0$). 
The branching ratio 
$$\tilde{\Gamma}_{f_1\to\gamma\gamma}\times \Gamma_{f_1\to K\bar{K}\pi}/\Gamma_{f_1}\simeq 1.9\pm 0.4~\text{keV}$$ is averaged by PDG~\cite{ParticleDataGroup:2022pth} among the existing measurements from 1987 to 2007.  The latest value of 
\begin{equation}\label{Eq:f1gaga_width}
	\tilde{\Gamma}_{f_1\to\gamma\gamma}\times \Gamma_{f_1\to K\bar{K}\pi}/\Gamma_{f_1}=3.2~\text{keV},
\end{equation} 
was reported by the L3 Collaboration~\cite{L3:2007obw}, which are the data we are comparing to in more detail in this work. The normalized TFF $F_{f_1\to\gamma^*\gamma^*}^{TL}(0,0)$ can in principle be adjusted as a free parameter. However, as discussed in subsection~\ref{subsec:IIIA}, an extra global normalization factor is introduced to describe the L3 events of $\gamma\gamma^*\to K_S^0 K^\pm \pi^\mp$. To avoid a large correlation in both parameters, we fix the TFF normalization to the L3 value of Eq.~\eqref{Eq:f1gaga_width} as $F_{f_1\to\gamma^*\gamma^*}^{TL}(0,0)=0.401$.

\subsection{Non-resonant channels for $\gamma\gamma^*\to K\bar{K}^*$}
In addition to the resonance production mechanism described above, we also need to 
incorporate the contributions from the contact term and the $t$- and $u$-channel charged $K$, $K^*$ exchange mechanisms, depicted in Fig.~\ref{Fig:diags}-(c,d,e). The pertinent interaction vertices for the charged $K$ and $K^*$ are described by the effective Lagrangians below: 
\begin{align}
  \mathcal{L}_{\gamma KK} &= i e A_\mu (K^+ \partial^\mu K^- - K^-\partial^\mu K^+), \nonumber\\	
  \mathcal{L}_{\gamma K^*K^*} & = -i e \biggl[ F^{\mu\nu} K^{*-}_\mu K^{*+}_\nu  \nonumber\\
  &\qquad + A_\mu  (K^{*+})^\nu \bigl( \partial^\mu (K^{*-})_\nu - \partial_\nu (K^{*-})^\mu\bigr) \nonumber \\
  &\qquad -  A_\mu  (K^{*-})^\nu \bigl(\partial^\mu (K^{*+})_\nu - \partial_\nu (K^{*+})^\mu\bigr) \biggr], \nonumber\\	
  \mathcal{L}_{\gamma K K^*} &= \frac{e g_{\gamma KK^*}}{m_K} \epsilon^{\mu\nu\alpha\beta} F_{\mu\nu} \nonumber \\
  &\quad \times \biggl[(K^{*+})_\alpha \partial_\beta K^- + (K^{*-})_\alpha \partial_\beta K^+\biggr],	\nonumber \\
  \mathcal{L}_{\gamma\gamma K K^*} &= - i \frac{e^2 g_{\gamma KK^*}}{m_K} \epsilon^{\mu\nu\alpha\beta} F_{\mu\nu} \nonumber \\
  &\quad \times \biggl[ (K^{*+})_\alpha K^- - (K^{*-})_\alpha K^+ \biggr] A_\beta,
\end{align}
with the electromagnetic tensor $F_{\mu\nu}= \partial_\mu A_\nu - \partial_\nu A_\mu$. For the spin-1 $K^*$ fields, we use the triple gauge boson interaction terms as in the $\mathrm{SU(2)}\otimes\mathrm{U(1)}_y$ Yang-Mills theory, which indicates the non-minimal term $F^{\mu\nu} K^{*-}_\mu K^{*+}_\nu$ in $\mathcal{L}_{\gamma K^*K^*}$. Such effective interaction guarantees tree-level unitarity for the non-resonant process in contrast to minimal substitution. The dimensionless coupling $g_{\gamma KK^*}$ is determined from the decay width of $K^{*+}\to K^{+}\gamma$, 
\begin{equation}
 \Gamma_{K^{*+}\to K^+\gamma} = \frac{e^2g_{\gamma KK^*}^2}{3\pi m_{K}^2} 	\left[q_{K^{*+}\to K^+\gamma}(M_{K^*}^2)\right]^3,
\end{equation}
with the rest frame momentum of the $K^{*+}\to K^+\gamma$ channel 
\begin{equation}
  	q_{K^{*+}\to K^+\gamma}(s)= \frac{\lambda^{1/2}(s, m_{K}^2, 0)} {2\sqrt{s}}.
\end{equation}
Using the experimental value of $\Gamma_{K^{*+}\to K^+\gamma}=50\pm 5$~keV given in PDG~\cite{ParticleDataGroup:2022pth}, one obtains $g_{\gamma KK^*}=0.203$. 

The amplitudes for the $\gamma(q_1)\gamma^*(q_2) \to \bar{K}^*(p_1) K(p_2)$ reaction corresponding to Fig.~\ref{Fig:diags}-(c,d,e) are expressed as follows:  
\begin{equation}\label{Eq:nonresAmp}
\begin{aligned}
   \mathcal{M}_\text{non-res.} =& \varepsilon_{\mu}(q_1,\lambda_1)\varepsilon_{\nu}(q_2,\lambda_2)	\varepsilon_{\alpha}^*(p_1,\Lambda_{K^*}) \\
   &\times \biggl(  \mathcal{M}_{(c)}^{\mu\nu\alpha} +  \mathcal{M}_{(d)}^{\mu\nu\alpha} +  \mathcal{M}_{(e)}^{\mu\nu\alpha}\biggr),
\end{aligned}
\end{equation}
with 
\begin{align}
  \mathcal{M}_{(c)}^{\mu\nu\alpha} & = - \frac{2 \, e^2\,g_{\gamma KK^*}}{m_K}\epsilon^{\mu\nu\alpha\beta} \\
  & \quad \times \biggl[ F_K(q_2^2)\,  {(q_1)}_\beta  - F_{\gamma^* KK^*} (q_2^2)\, {(q_2)}_\beta \biggr], \nonumber\\
  \mathcal{M}_{(d)}^{\mu\nu\alpha} &= \frac{2 \, e^2}{m_K} \, g_{\gamma KK^*}\Biggl[\frac{\epsilon^{\mu\sigma\alpha\beta} {(q_1)}_\sigma {(p_1)}_\beta (2p_2-q_2)^\nu}{q_2^2 - 2q_2\cdot p_2} F_{K}(q_2^2) 	\nonumber\\
  & \quad +  \frac{\epsilon^{\nu\sigma\alpha\beta} {(q_2)}_\sigma {(p_1)}_\beta (2p_2-q_1)^\mu}{q_1^2 - 2q_1\cdot p_2}  F_{\gamma^* KK^*}(q_2^2) \Biggr], \nonumber\\
  \mathcal{M}_{(e)}^{\mu\nu\alpha} &= \frac{2\, e^2\,g_{\gamma KK^*}}{m_K}\Biggl[ \frac{\epsilon^{\nu\sigma \rho \beta} {(q_2)}_\sigma {(p_2)}_\beta}{q_1^2 - 2 q_1\cdot p_1} F_{\gamma^* KK^*}(q_2^2)   \nonumber\\
  &\times \biggl( g_{\rho}^{~\alpha} (2p_1-q_1)^\mu - g_{\rho}^{~\mu}(p_1-2q_1)^\alpha - g^{\mu\alpha}{(p_1+q_1)}_\rho\biggr) \nonumber \\
  & + \frac{\epsilon^{\mu\sigma \rho \beta} {(q_1)}_\sigma {(p_2)}_\beta}{q_2^2 - 2 q_2\cdot p_1}F_{K}(q_2^2)  \biggl( g_{\rho}^{~\alpha} (2p_1-q_2)^\nu \nonumber\\
  &\quad - g_{\rho}^{~\nu}(p_1-2q_2)^\alpha - g^{\nu\alpha}{(p_1+q_2)}_\rho\biggr)  \Biggr],
\end{align}	
where we included the electromagnetic kaon form factor $F_K(q_2^2)$ and the vector meson transition form factor $F_{\gamma^* KK^*}(q_2^2)$. Both are considered in the monopole form, 
\begin{equation}
\begin{aligned}
   F_K(q_2^2) &= \frac{1}{1+Q_2^2/\Lambda_K^2},\\
   F_{\gamma^* KK^*}(q_2^2) &= \frac{1}{1+Q_2^2/\Lambda_{K^*}^2}, 	
\end{aligned}
\end{equation}
with the monopole masses $\Lambda_K=872$ MeV~\cite{Danilkin:2018qfn} and $\Lambda_{K^*}=M_{K^*}=893.5$ MeV. The electromagnetic form factor $F_{K^*}(q_2^2)$ is set equal to the $F_{K}(q_2^2)$ in the above amplitude to satisfy gauge invariance. 

To extend the above amplitude into the high-energy region, as we used in Ref.~\cite{Ren:2022fhp}, we first express the amplitude using the Lorentz tensor decomposition, 
\begin{equation}
    \mathcal{M}_{(c+d+e)}^{\mu\nu,\alpha}=\sum\limits_{i=1}^9 T_i^{\mu\nu,\alpha}(q_1,q_2;p_1-p_2)\, F_i(W^2,t,u).
\end{equation}
Here $T^i_{\mu\nu,\alpha}$ stands for the complete set of the nine gauge invariant tensors for the $\gamma\gamma^*\to VP$ reaction, and $F_i$ corresponds to the scalar functions. Both expressions can be found in the Appendix \ref{App}. 

We assume that in the high energy region the above amplitude is dominated by Regge poles. Then, one can calculate the residues of the $K$ and $K^*$ Regge exchange based on the amplitudes calculated with Feynman propagators. We apply the Regge trajectory of $K$ and $K^*$ to replace the corresponding propagators. 
The Reggeized propagators of the $K$ and $K^*$ mesons in the $t$-channel are
\begin{equation}
\begin{aligned}
\frac{1}{t-m_K^2} \to \mathcal{P}^K(W^2, t) &\equiv \left(\frac{W^2}{W_0^2}\right)^{\alpha_K(t)} \frac{\pi \alpha_K^{\prime}}{\sin \left(\pi \alpha_K(t)\right)} \\
&\quad \times \left(\frac{1+e^{-i \pi \alpha_K(t)}}{2 \Gamma\left(1+\alpha_K(t)\right)}\right),\\
\frac{1}{t-M_{K^*}^2}  \to \mathcal{P}^{K^*}(W^2, t) & \equiv\left(\frac{W^2}{W_0^2}\right)^{\alpha_{K^*}(t)-1} \frac{\pi \alpha_{K^*}^{\prime}}{\sin \left(\pi \alpha_{K^*}(t)\right)} \\
&\quad \times \left(\frac{-1+e^{-i \pi \alpha_{K^*}(t)}}{2 \Gamma\left(\alpha_{K^*}(t)\right)}\right),
\end{aligned}
\end{equation}
with the trajectories of $K$ and $K^*$ mesons are given by $\alpha_K(t) = 0.7 (t-m_K^2)$ and $\alpha_{K^*}(t) = 0.25 + 0.83 t$~\cite{Guidal:1997hy}, respectively. The same form applies to the $K$ and $K^*$ Reggeized propagators in the $u$-channel. The Gamma function $\Gamma(\alpha(t))$ ensures that the propagator only has poles in the timelike region. The mass scale is conventionally taken as $W_0=1 ~\mathrm{GeV}$. 

\subsection{Cross sections of $\gamma\gamma^*\to K\bar{K}^*$ reaction}

Since the $\gamma\gamma^*\to K\bar{K}^*$ process enters the cross section for the unpolarized single tagged $e^+e^-\to e^+e^-K\bar{K}^*$ process, we first present the differential cross section for the latter reaction 
\begin{equation}\label{Eq:eeXsec}
\begin{aligned}
\frac{d \sigma}{d Q_2^2 \, d W^2} = \frac{\tilde{F}_2^{++}}{Q_2^2\left(W^2+Q_2^2\right)} &\bigl[\sigma_{TT}(W^2,0,Q_2^2) \\
& +\varepsilon \,\sigma_{TL}(W^2,0,Q_2^2)\bigr],
\end{aligned}
\end{equation}
in order to fix the convention for $\sigma_{TT}$ and $\sigma_{TL}$. Here, $W^2=(q_1+q_2)^2$ denotes the invariant mass of the $K\bar{K}^*$ system, the dimensionless quantity $\tilde{F}_2^{++}$ stands for the integrated transverse virtual photon flux factor, and $\varepsilon$ is the longitudinal photon polarization parameter. Their expressions and the details of the derivation of Eq.~\eqref{Eq:eeXsec} are given in Appendix~\ref{App:eeXsec}.

The differential polarized cross sections of the $\gamma(q_1)\gamma^*(q_2)\to \bar{K}^*(p_1)K(p_2)$ process are then given by 
\begin{equation}\label{Eq:diffXsec}
\begin{aligned}
    \frac{d \sigma_{TT}}{d\cos\theta} & =  \frac{1}{64 \pi W^2} \frac{\lambda^{1/2}(W^2, M_{k^*}^2, m_K^2)}{\lambda^{1/2}(W^2, 0, -Q_2^2)} \\
    &\quad \times \sum\limits_{\Lambda_{K^*}}\bigl(|\mathcal{M}_{++,\Lambda_{K^*}}|^2 + |\mathcal{M}_{+-,\Lambda_{K^*}}|^2\bigr), \\
    \frac{d\sigma_{TL}}{d\cos\theta} & =  \frac{1}{32 \pi W^2} \frac{\lambda^{1/2}(W^2, M_{k^*}^2, m_K^2)}{\lambda^{1/2}(W^2, 0, -Q_2^2)}  \sum\limits_{\Lambda_{K^*}} |\mathcal{M}_{+0,\Lambda_{K^*}}|^2,
\end{aligned}
\end{equation}
where  
$\mathcal{M}_{\lambda_1\lambda_2,\Lambda_{K^*}}$ stands for the helicity amplitude of the photon fusion process, and
$\theta$ is the scattering angle in the $\gamma\gamma^*$ c.m. system.  

\begin{figure*}[t]
    \centering
    \includegraphics[width=0.7\textwidth]{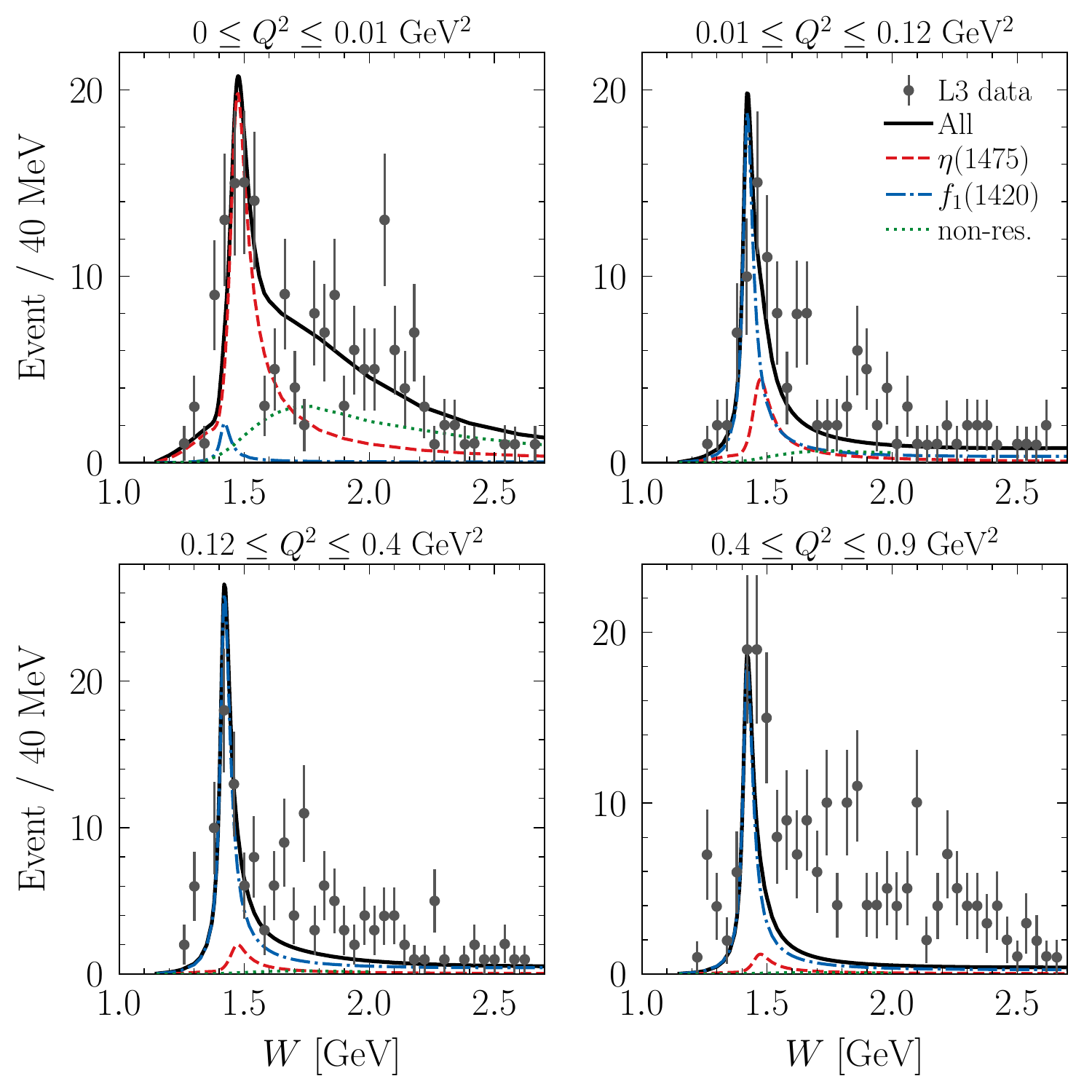}
    \caption{The $W=M(K_S^0K^\pm \pi^\mp)$ dependence for the events of $e^+e^- \to e^+e^- K_S^0 K^\pm \pi^\mp$ reaction via the photon-photon fusion for the four $Q^2$ bins: $0-0.01$ GeV$^2$, $0.01-0.12$ GeV$^2$, $0.12-0.4$ GeV$^2$, and  $0.4-0.9$ GeV$^2$. The gray dots are the data points from the L3 Collaboration~\cite{L3:2007obw}. The black solid curve denotes the full results of our model, and the red dashed, blue dot-dashed, and green dotted curves are the contributions of $\eta(1475)$, $f_1(1420)$, and non-resonant channels, respectively.}
    \label{Fig:L3Event}
\end{figure*}

Before proceeding to the numerical results, it is necessary to discuss the relative phases of the  $\eta(1475)$, $f_1(1420)$, and non-resonant contributions. For the $\eta(1475)$ contribution, only helicity amplitudes  with $\lambda_{1,2}=1$ and $\Lambda_{K^*}=0$ are non-zero. For the $f_1(1420)$ contribution, the non-zero helicity amplitudes occur when the relative helicity of both photons is either zero or one. We found an interesting result: the constructive and destructive interferences between the $\eta(1475)$ and $f_1(1420)$ amplitudes do not affect the polarized cross sections for $\gamma\gamma^*\to K\bar{K}^*$. This arises from the fact that $\mathcal{M}_{++,0}(\bar{\eta})$ is independent of $\cos\theta$, while $\mathcal{M}_{++,0}(f_1)$ is proportional to $\cos\theta$. Consequently, the differential cross section $d\sigma_{++}/d\cos\theta$ with constructive/destructive interference modes $M_{++,0}(\bar{\eta})\pm \mathcal{M}_{++,0}(f_1)$ are symmetric in $\cos\theta$ ranging from $-1$ to $1$. This symmetry leads to the same total cross section $\sigma_{TT}$ upon integration over $\cos\theta$. 
Thus, the experimental measurement of $d\sigma_{TT}/d\cos\theta$ allows to distinguish the relative phase between the $\eta(1475)$ and $f_1(1420)$ contributions. Including the non-resonant contribution, we find that the constructive interference with the $\eta(1475)$ and $f_1(1420)$ components yields a better global description of the L3 data compared to the ones with the destructive interference, as shown in Fig.~\ref{Fig:L3Event}.

\begin{table}[t]
\caption{Values of resonance $(R)$ parameters used in our model.}	
\begin{tabular}{c|c|c|c|c}
\hline \hline 
    & $m_R\,[\mathrm{MeV}]$ & $\Gamma_R\,[\mathrm{MeV}]$ & coupling & TFF  \\[0.5em]
\hline 
& &  & $g_{\bar{\eta}KK^*}$ & $F_{\bar{\eta} \gamma^* \gamma^*}(0,0) $ \\[0.5em]
$\eta(1475)$ & 1475 & 90 & 2.04 & 0.0414\, GeV$^{-1}$ \\
\hline 
& & & $g_{f_1 KK^*}$ & $F^{TL}_{f_1 \gamma^* \gamma^*}(0,0)$  \\[0.5em]
$f_1(1420)$ & 1426.3 & 54.5  & 1.027 & 0.401\\
\hline
& & & $g_{\gamma KK^*}$ &   \\[0.5em]
$K^*(892)$ & 893.5 & 51.4 & 0.203 &  \\
\hline \hline
\end{tabular}
\label{Tab:parameters}
\end{table}

\section{Results and discussion}\label{SecIII}
In this section, we first describe the available experimental data related to the $\gamma\gamma^*\to K\bar{K}^*(892)$ process, i.e. the L3 measurement of the $e^+e^-\to  e^+e^- \gamma\gamma^*\to  e^+e^- K_S^0 K^\pm \pi^\mp$ events~\cite{L3:2007obw}, to constrain the TFFs of $f_1(1420)$ state. Then, the theoretical prediction of the polarized cross sections of $\gamma\gamma^*\to K\bar{K}^*(892)$ reaction is presented for the forthcoming BESIII measurement. 

\subsection{Description of the L3 data}\label{subsec:IIIA}
In our phenomenological model, the free parameter is the dipole mass $\Lambda_{f_1}$, which enters the transition form factor~Eq.~\eqref{Eq:TFFf1} for the $f_1(1420)$ resonance. The other parameters, along with the PDG values of masses and widths of resonances~\cite{ParticleDataGroup:2022pth}, are listed in~Table~\ref{Tab:parameters}. The L3 data of $\gamma\gamma^*\to K_S^0 K^\pm \pi^\mp$ reaction, where the $f_1(1420)$ state is prominently observed in  five  $Q^2$ bins from $0$ to $7$ GeV$^2$, provide a way to empirically determine $\Lambda_{f_1}$. 

To apply our cross section model to the L3 measurement,  
we need to establish the connection between the polarized cross sections of the $\gamma\gamma^*\to K\bar{K}^*$ reaction and the events of the $\gamma\gamma^*\to K_S^0K^\pm \pi^\mp$ process for $e^+e^-$ c.m. energies of $183\sim 209$ GeV as reported by the L3 Collaboration.
First, we take into account the $K^{*-} \to (K\pi)^-$ decay effectively by using 
\begin{equation}\label{Eq:Xsec_KKpi}
\begin{aligned}
    \sigma_{\gamma\gamma^*\to K^+(K\pi)^-}(W^2) &= \int_{(m_K+m_\pi)^2}^{(W-m_K)^2} d W_{K^*}^2\, F(W_{K^*}^2) \\
    &\quad \times \sigma_{\gamma\gamma^*\to K\bar{K}^*}(W^2, W_{K^*}), 
\end{aligned}
\end{equation}
to obtain the unpolarized cross section of $\gamma\gamma^*\to KK\pi$ reaction, where the $K^*$ mass in the $\sigma_{\gamma\gamma^*\to K\bar{K}^*}$ is replaced by the variable $W_{K^*}$, starting from the $K\pi$ threshold.
The lineshape function $F(W_{K^*}^2)$ is defined as 
\begin{equation}~\label{Eq:lineshape}
  F(W_{K^*}^2) = \frac{1}{\pi} \frac{M_{K^*}\,\Gamma_{K^*}(W_{K^*}^2)}{(W_{K^*}^2-M_{K^*}^2)^2 + \bigl(M_{K^*}\Gamma_{K^*}(W_{K^*}^2)\bigr)^2}, 
\end{equation}
in order to satisfy the normalization 
\begin{equation}
	\int_{(m_K+m_\pi)^2}^\infty dW_{K^*}^2 F(W_{K^*}^2)=1,
\end{equation} where the energy-dependent width in Eq.~\eqref{Eq:lineshape} is given by 
\begin{equation}
    \Gamma_{K^*}(W_{K^*}^2) = \Gamma_{K^*}(M_{K^*}^2) \sqrt{\frac{W_{K^*}^2 - (m_K+m_\pi)^2}{M_{K^*}^2 - (m_K+m_\pi)^2}}.
\end{equation}
To access the unpolarized cross section combination $\sigma_{\gamma\gamma^*\to K\bar{K}^*}=\sigma_{TT} + \varepsilon \, \sigma_{TL}$, which enters Eq.~\eqref{Eq:eeXsec}, one needs to know the longitudinal photon polarization parameter $\varepsilon$. In principle, such value is provided by the L3 experimental conditions. Since the $e^+e^-$ beam energy (from $183$ to $209$ GeV) at LEP is significantly larger than the $\gamma\gamma^*$ total energy $W=1\sim 3$ GeV, one can safely approximate $\varepsilon\approx 1$.

We obtain the unpolarized cross section of the $\gamma\gamma^*\to K^+(K\pi)^-$ by allowing $\bar{K}^*$ decay to $K\pi$. More specifically, we have this process: $\gamma\gamma^*\to K^+K^{*-}\to K^+ (K\pi)^-$. In order to estimate the $\gamma\gamma^*\to K^+ \bar{K}^0\pi^-$ reaction, its cross section is given by 
\begin{equation}
	\sigma_{\gamma\gamma^*\to K^+\bar{K}^0\pi^-}= \frac{2}{3} \,\sigma_{\gamma\gamma^*\to K^+(K\pi)^-}.
\end{equation}
However, the L3 data are for the $\gamma\gamma^*\to K_{S}^0 K^{\pm}\pi^{\mp}$ process and include two intermediate processes $\gamma\gamma^* \to K_S^0(K \pi)^0$ and $\gamma\gamma^* \to (K_S^0\pi^\mp) K^\pm$. The charged $KK^*$ process, i.e. $\gamma\gamma^* \to (K_S^0\pi^\mp) K^\pm$, can be directly related to the $\gamma\gamma^*\to K^+\bar{K}^0\pi$ process studied above under isospin conservation. Ignoring the tiny effect of CP violation, we have $|K_{S}^0\rangle = \frac{1}{\sqrt{2}}(|K^0\rangle + |\bar{K}^0\rangle)$, then the L3 measured cross section can be related to the cross section of $\gamma\gamma^*\to K^+ \bar{K}^0\pi^-$ via 
 \begin{equation}
 	\sigma_{\gamma\gamma^*\to (K_{S}^0\pi^-) K^+}= \frac{1}{2}\,\sigma_{\gamma\gamma^*\to K^+\bar{K}^0\pi^-}.
 \end{equation}
Thus, we have the following relation 
\begin{equation}
		\sigma_{\gamma\gamma^*\to (K_{S}^0\pi^-) K^+} =  \frac{1}{3} \,\sigma_{\gamma\gamma^*\to K^+(K\pi)^-}.
\end{equation}
Another intermediate process of L3 measurement, $\gamma\gamma^* \to K_S^0(K \pi)^0$ involves the neutral $K$ and $K^*$. Since both $\eta(1475)$ and $f_1(1420)$ have isospin zero, the  $\gamma\gamma^* \to K^0\bar{K}^{*0}$ amplitudes are identical to those for $\gamma\gamma^* \to K^+ K^{*-}$, given in Eqs.~\eqref{Eq:AmpEta} and \eqref{Eq:Ampf1}, respectively.  However, for the non-resonant channel, the contribution to $\gamma\gamma^*\to K^0\bar{K}^{*0}$ may differ from that of $\gamma\gamma^*\to K^+K^{*-}$ as the charged $K$ and $K^*$ exchange processes of Figs.~\ref{Fig:diags}(c-e) do not contribute for the former process, and require further investigation. As the non-resonant process is quite small in the kinematic region shown in this work, we assume in the comparison with the L3 data that they are the same for both channels.
Finally, considering both intermediate processes of the L3 experiment, we establish the following relation between the cross sections 
\begin{equation}
		\sigma_{\gamma\gamma^*\to K_{S}^0 K^{\pm}\pi^{\mp} } =  \frac{2}{3} \,\sigma_{\gamma\gamma^*\to K^+(K\pi)^-}.
\end{equation}

Next, we need to establish the relation between our $\gamma\gamma^*$ cross section prediction and the L3 events for the $e^+ e^- \to e^+e^- K_S^0 K^\pm \pi^\mp$ process through the two-photon collision. Such comparison typically requires specific information about the L3 detector, such as the virtual photon flux factors, the minimal and maximal virtualities of photons, etc, as presented in Appendix~\ref{App:eeXsec}. Based on the differential cross section~given in Eq.~\eqref{Eq:eeXsec}, we parameterize the experimental events as measured by L3 as: 
\begin{equation}
    \mathrm{Events}(W^2) = \mathcal{N}\int_{Q^2_\mathrm{low}}^{Q^2_\mathrm{high}} dQ^2\, \frac{1}{Q^2}\, \frac{2}{3}\sigma_{\gamma\gamma^*\to K^+(K\pi)^-}(W^2),
\end{equation}
where the integral limits correspond to the different $Q^2$ bins. The normalization factor $\mathcal{N}$ depends on the specific details of the experimental detector. Since this information is not accessible, we approximate a global  normalization factor in practice. This factor is determined to simultaneously describe the L3 events across all $Q^2$ bins: $Q^2\sim [0,0.01]$, $[0.01-0.12]$, $[0.12-0.4]$, $[0.4-0.9]$, and $[0.9-7]$ GeV$^2$. Note that the existence of such a single normalization factor leads to a meaningful comparison with the L3 data.

\begin{figure*}[t]
    \centering
    \includegraphics[width=0.48\textwidth]{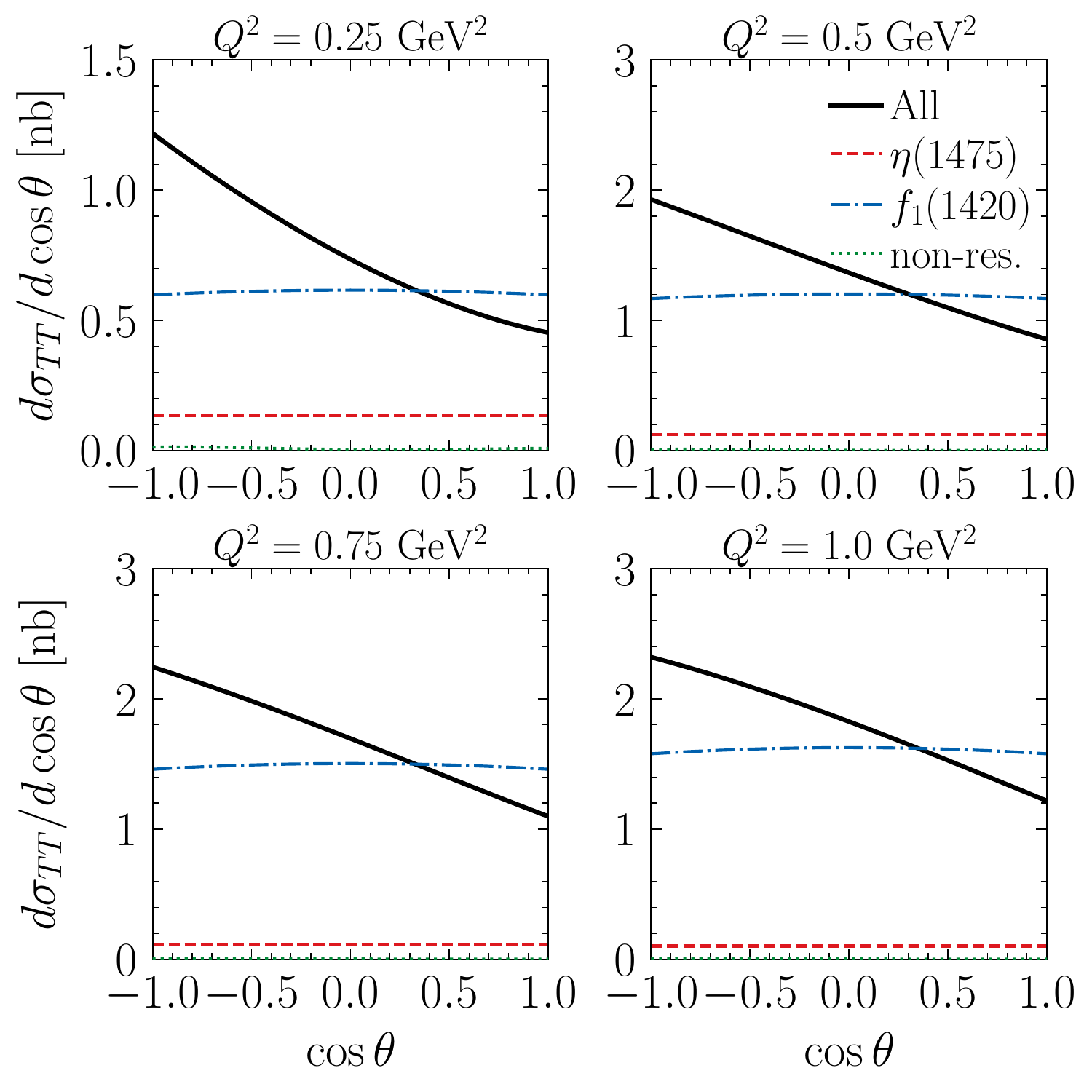} ~~~~
    \includegraphics[width=0.48\textwidth]{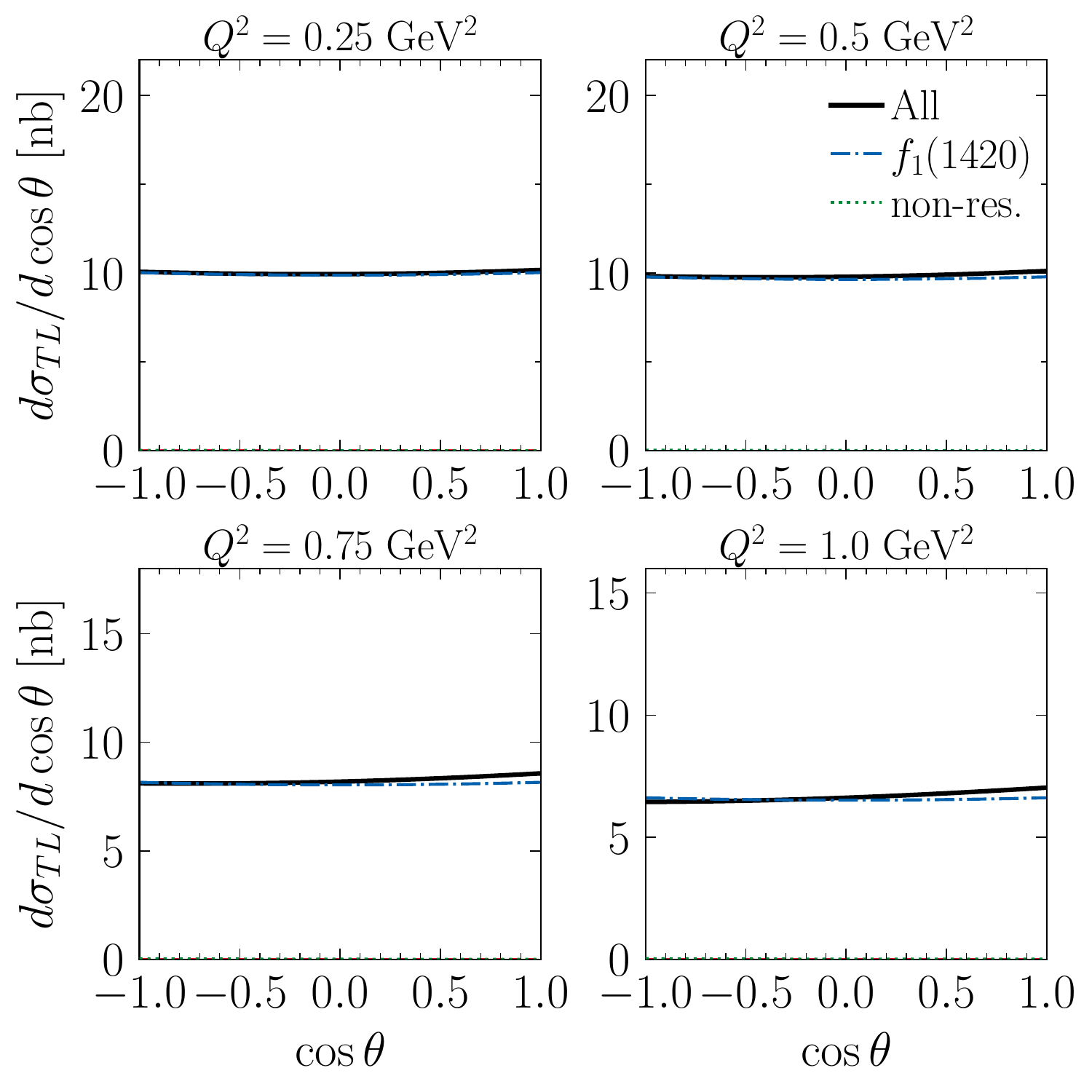} 
    \caption{Prediction for the differential cross sections $d\sigma_{TT}/d\cos\theta$ (left panels) and $d\sigma_{TL}/d\cos\theta$ (right panels)  for $\gamma\gamma^*\to K^\pm K^{*\mp}$ with $Q^2=0.25$, $0.5$, $0.75$, $1.0$ GeV$^2$ for $W = 1.42$~GeV. The curve notations are the same as in Fig.~\ref{Fig:L3Event}.}
    \label{Fig:dsigdcos_BESIII}
\end{figure*}

\begin{figure*}[t]
    \centering
    \includegraphics[width=0.48\textwidth]{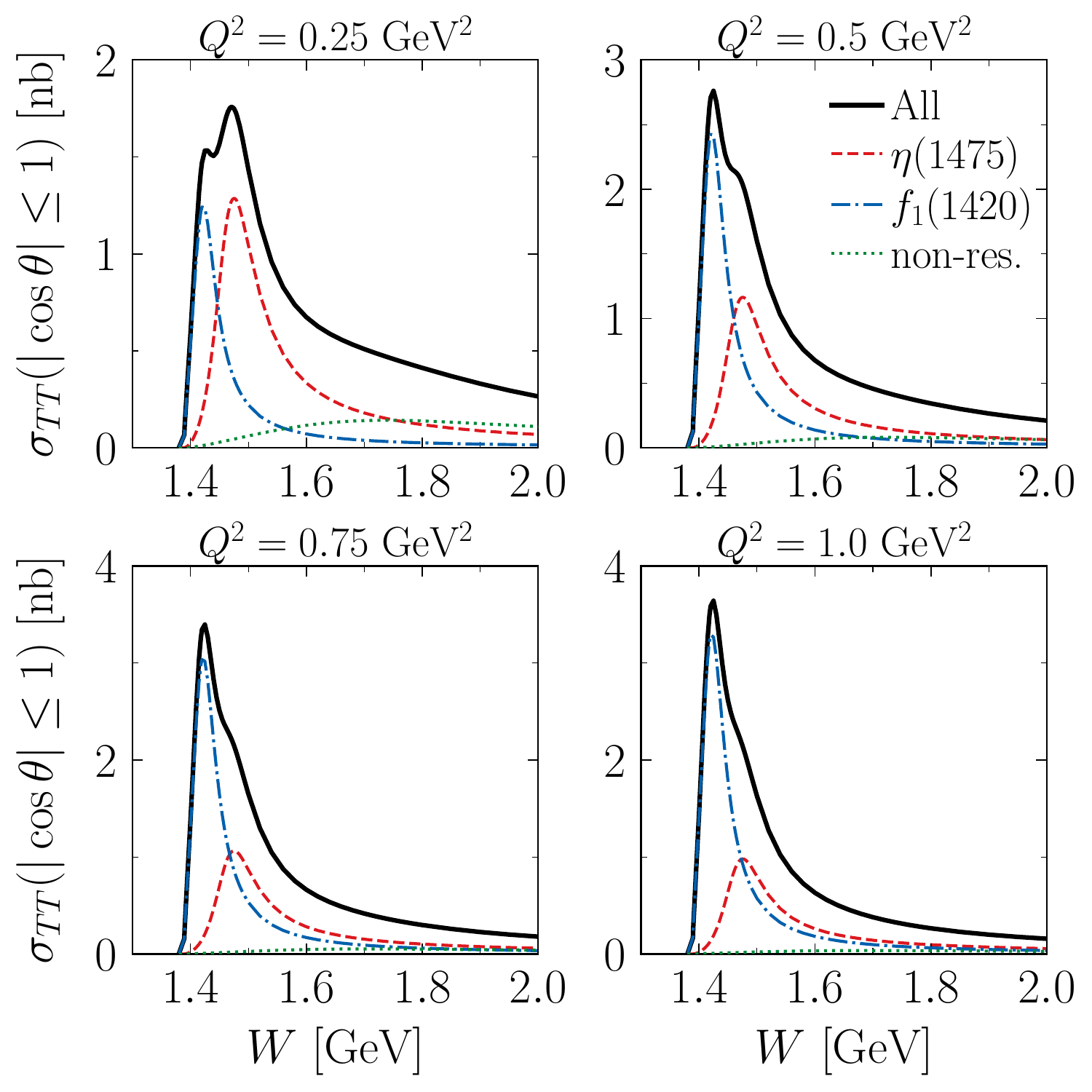} ~~~~
    \includegraphics[width=0.48\textwidth]{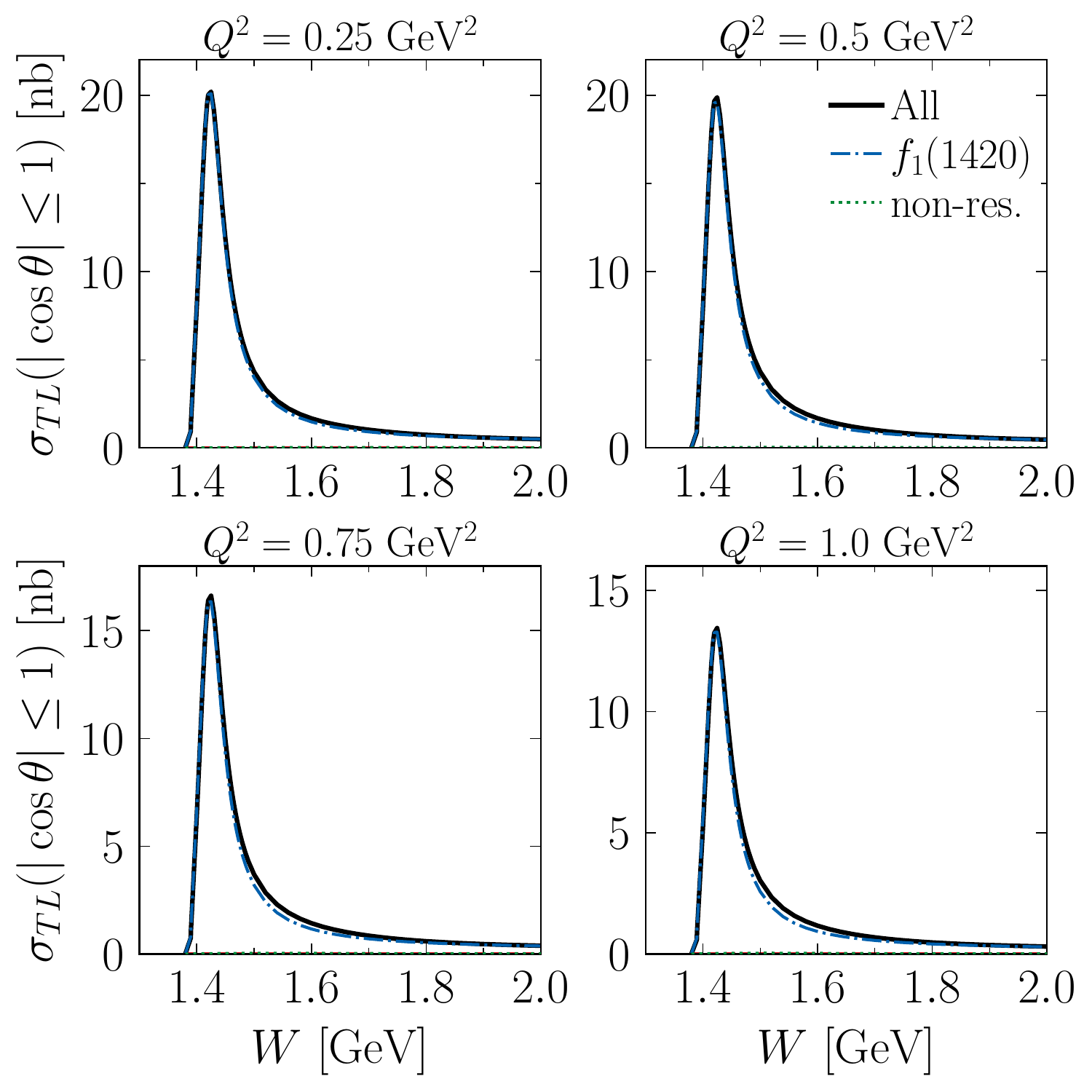} 
    \caption{Prediction for $\sigma_{TT}$ (left panels) and $\sigma_{TL}$ (right panels) cross sections for $\gamma\gamma^*\to K^\pm K^{*\mp}$ with $Q^2=0.25$, $0.5$, $0.75$, $1.0$ GeV$^2$ and with the full angular coverage $|\cos\theta|\leq 1$. The curve notations are the same as in Fig.~\ref{Fig:L3Event}.}
    \label{Fig:sig_BESIII}
\end{figure*}

A reasonable description of all events, as shown in Fig.~\ref{Fig:L3Event}~\footnote{Note that the comparison with the broad and high $Q^2\sim[0.9,7]$ GeV$^2$ bin is not shown as it is beyond the applicability of our model, but a reasonable description of the L3 data is still achieved.}, can be achieved by adjusting the dipole mass scale in the $f_1(1420)$ TFF of Eq.~\eqref{Eq:TFFf1} to $\Lambda_{f_1}=920$ MeV. This value is consistent with the one given in Ref.~\cite{L3:2007obw}. From Fig.~\ref{Fig:L3Event}, we observe that the $\eta(1475)$ contribution is dominant for quasi-real photons and gradually decreases with increasing photon virtuality. In the low $Q^2$ region~$[0-0.01]$ GeV$^2$, the contribution of the non-resonant channel is large relative to the $f_1(1420)$ result. This is because the production of axial-vector mesons at any small $Q^2$ is suppressed. Furthermore, the constructive interference between the $\eta(1475)$ and the non-resonant channels provides  a rather good description of the L3 events at large $\gamma\gamma^*$ c.m. energy. For the $Q^2$ range of $0.01-0.12$ GeV$^2$and larger, the $f_1(1420)$ contribution is dominant. Thus, the theoretical calculation of the $e^+e^- \to e^+e^- K_S^0 K^\pm \pi^\mp$ process is sensitive to the $f_1(1420)$ TFF. 

In the last two $Q^2$ bins, we notice that the L3 data show an enhancement relative to our results at very low $\gamma\gamma^*$ energy. This discrepancy arises due to the missing contribution of the $f_1(1285)$ resonance in our model. Additionally, the deviation from the L3 data is also observed on the higher energy side of the $f_1(1420)$ resonance. 
One reason is that we use quasi-two-body states to mimic the three-body final states, neglecting the non-resonant mechanisms involving $KK\pi$ states in our analysis. Such contributions have been noticed in the L3 data analysis, e.g. the background contribution to the $K^*$ invariant mass distribution, as shown in~Fig.~7 of Ref.~\cite{L3:2007obw}.

To further improve upon the description of the L3 data, it is possible to include the contributions from the $f_1(1285)$ exchange in the $s$-channel and potentially higher mass resonances like $\eta(1760)$~\cite{Belle:2012uhr}. However, including these higher resonances in our model is not expected to significantly cahnge the data description within the energy region of $f_1(1420)$. The reason is twofold: these higher resonances are far away from the $f_1(1420)$ state; the total cross section of $\gamma\gamma^*\to K\bar{K}^*$ is dominated by $\sigma_{TL}$, while $\eta(1760)$ can only contribute to the transverse-transverse part. For a more realistic description, we plan to extend the current model of the $\gamma\gamma^*\to K\bar{K}^*$ reaction to the process with actual three-body final states via the $K^*$ decay to $K\pi$, i.e. $\gamma\gamma^*\to K^\pm K^{*\mp}\to K^+K^-\pi^0$, which is an ongoing analysis process at BESIII. 

\subsection{Prediction of polarized $\gamma\gamma^*\to K^\pm K^{*\mp}(892)$ cross sections}

Based on the reasonable description of the L3 data in the $f_1(1420)$ region, we first present our predictions for the polarized differential cross sections in Fig.~\ref{Fig:dsigdcos_BESIII} with $W=M_{f_1}$. We take $Q^2=0.25$, $0.5$, $0.75$, $1.0$ GeV$^2$ to cover the range of forthcoming BESIII data for the $e^+e^- \to e^+e^-K^+K^-\pi^0$ process. Since both $K$ and $K^*$ are charged in the process measured by BESIII, our model for the non-resonant process is directly applicable.
From Fig.~~\ref{Fig:dsigdcos_BESIII}, we notice that the constructive interference between the $S$-wave $f_1(1420)$ and the $P$-wave $\eta(1475)$ channels, determined by the L3 data, is clearly shown in the transverse part. The contribution from the non-resonant channel is rather small. In contrast, the $f_1(1420)$ production mechanism dominates the $d\sigma_{TL}/d\cos\theta$ cross section, which leads to the quasi-angular-independent results across all $Q^2$ values. The forthcoming BESIII data will provide valuable validation of these predictions.

Using the phenomenological model of $\gamma\gamma^*\to K^\pm K^{*\mp}(892)$, we {\color{blue}also} predict the polarized cross sections, $\sigma_{TT}$ and $\sigma_{TL}$, in Fig.~\ref{Fig:sig_BESIII}. 
For the transverse cross section $\sigma_{TT}$ at $Q^2=0.25$ GeV$^2$, we found a broad peak around $\sqrt{s}=1.45$ GeV, consistent with a preliminary analysis of the BESIII data~\cite{Effenberger_MasterThesis}. This peak originates from the interference of the two closely located resonances, $\eta(1475)$ and $f_1(1420)$. Our study provides a natural explanation, which could facilitate the extraction of the $f_1(1420)$ resonance parameters from experimental data. Furthermore, the constructive interference between the $\eta(1475)$ production channel and the non-resonant channel can be further validated by the forthcoming high-statistics BESIII data. As $Q^2$ increases, one notices from Fig.~\ref{Fig:sig_BESIII} that the contribution from  the $f_1(1420)$ production mechanism dominates. Accordingly, the contributions from the $\eta(1475)$ and the non-resonant channels decrease. However, a small shoulder on the higher energy side of the $f_1(1420)$ peak remains visible. 

Since the $\eta(1475)$ channel does not contribute to the $\sigma_{TL}$ cross section, the $f_1(1420)$ production mechanism dominates $\sigma_{TL}$ for all $Q^2$ values  shown and accounts for over $95\%$ of the total $\sigma_{TL}$. As $Q^2$ increases, the $f_1(1420)$ production cross section decreases gradually, attributed to the dipole form of the TFF. Thus, forthcoming BESIII data are promising to provide an extraction of the $f_1(1420)$ TFF. 
Furthermore, one can see that the magnitude of $\sigma_{TL}$ is approximately 10 times larger than that of $\sigma_{TT}$ in the low $Q^2$ region. This difference arises because $\sigma_{TT}$ is suppressed by $Q^2/(2M_{f_1}^2)$, as seen from Eq.~\eqref{Eq:XSec_TFFs} and discussed above.

\begin{figure*}[t]
    \centering
    \includegraphics[width=0.7\textwidth]{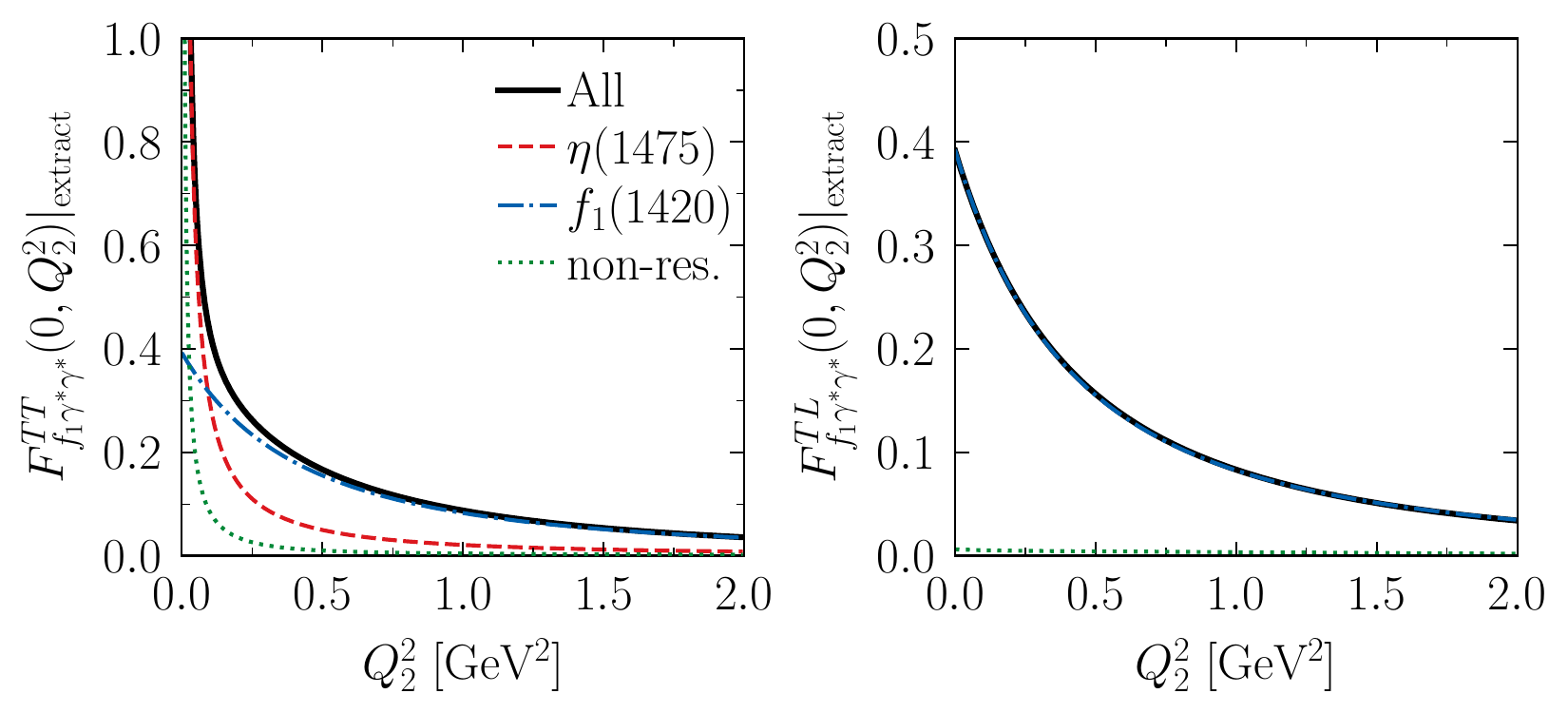}
    \caption{The extracted quantities $\left.F_{f_1\gamma^*\gamma^*}^{TT}\right|_\mathrm{extract}$ and $\left.F^{TL}_{f_1\gamma^*\gamma^*}\right|_\mathrm{extract}$ (defined through Eq.~\eqref{Eq:extractTFFs}) as functions of $Q_2^2$. The curve notations are the same as Fig.~\ref{Fig:L3Event}. The $f_1(1420)$ TFFs are given by the blue dot-dashed curves.}
    \label{Fig:TFFs}
\end{figure*}

Finally, from the measurement of the above-mentioned polarized cross sections, we demonstrate the possibility to extract the $F_{f_1\gamma^*\gamma^*}^{TT}(0,Q_2^2)$ and $F_{f_1\gamma^*\gamma^*}^{TL}(0,Q_2^2)$ TFFs. Based on Eq.~\eqref{Eq:XSec_TFFs}, one can define the quantities at the $f_1(1420)$ resonance  at $W^2=M_{f_1}^2$ as:
\begin{align}\label{Eq:extractTFFs}
	&\left.\bigl[ F_{f_1\gamma^*\gamma^*}^{TT}(0,Q_2^2)\bigr]^2\right|_\mathrm{extract} \equiv 
	\frac{M_{f_1}\, \Gamma_{f_1}}{4\pi^2\alpha^2\,\mathrm{Br}(f_1\to K\bar{K}^*)} \nonumber\\
	&\qquad \times \frac{2M_{f_1}^4}{Q_2^4(1+Q_2^2/M_{f_1}^2)}  \sigma_{TT}(W^2=M_{f_1}^2,0,Q_2^2), \\
	&\left.\bigl[ F_{f_1\gamma^*\gamma^*}^{TL}(0,Q_2^2)\bigr]^2\right|_\mathrm{extract} \equiv 
	\frac{M_{f_1}\, \Gamma_{f_1}}{4\pi^2\alpha^2\,\mathrm{Br}(f_1\to K\bar{K}^*)}  \nonumber\\
	&\qquad \times \frac{M_{f_1}^2}{Q_2^2(1+Q_2^2/M_{f_1}^2)}  \sigma_{TL}(W^2=M_{f_1}^2,0,Q_2^2). \nonumber 
\end{align}
Here we employed the polarized cross sections of $\gamma\gamma^*\to K\bar{K}^*(892)$ instead of those of $\gamma\gamma^*\to f_1(1420)$ in Eq.~\eqref{Eq:XSec_TFFs}, by introducing the branching ratio $\mathrm{Br}(f_1\to K\bar{K}^*)$ for charged final states. 
The obtained quantities are presented in Fig.~\ref{Fig:TFFs} as functions of $Q_2^2$. 
Due to the significant contributions from $\eta(1475)$ and non-resonant channels at low $Q^2$,  the extracted quantity $F^{TT}_{f_1\gamma^*\gamma^*}$ is larger than the dipole parametrization (Eq.~\eqref{Eq:TFFf1}) for the $f_1(1420)$ TFF, which is given by the blue curve in Fig.~\ref{Fig:TFFs}.  As $Q^2$ increases, e.g. $Q_2^2\geq 0.5$  GeV$^2$, the contribution of $f_1(1420)$ channel dominants over the $
\eta(1475)$ channel, as shown in Fig.~\ref{Fig:sig_BESIII}. Consequently,  $F^{TT}_{f_1\gamma^*\gamma^*}$ can be reliably extracted in the larger $Q_2^2$ region. The situation regarding $F^{TL}_{f_1\gamma^*\gamma^*}$ form factor is more straightforward. One can directly extract $F^{TL}_{f_1\gamma\gamma^*}$  from $\sigma_{TL}$ via Eq.~\eqref{Eq:extractTFFs} in the whole $Q_2^2$ region, since the $\eta(1475)$ channel is forbidden, and the non-resonant contribution is relatively small.

\section{Conclusion}\label{SecIV}
In this work, we have developed a phenomenological model for the  $\gamma\gamma^*\to K\bar{K}^*(892)$ reaction. Our model includes the production mechanism of  the $\eta(1475)$ and $f_1(1420)$ resonances in the $s$-channel. Additionally, we have parametrized the  non-resonant contribution using the charged $K$ and $K^*$ crossed-channel exchanges. By performing the Lorentz tensor decomposition of the $\gamma\gamma^*\to PV$ amplitude, we employ the Regge trajectories to replace the $K$ and $K^*$ propagators, ensuring a correct high-energy behavior.

In order to constrain the transition form factor of $\gamma\gamma^*\to f_1(1420)$ in our model, we utilize the available L3 data from the $\gamma\gamma^*\to K_S^0 K^\pm \pi^\mp$ process, which leads to a dipole mass parameter $\Lambda_{f_1}=920$ MeV in good agreement with the L3 extraction. Subsequently, we predict the polarized cross sections within the $Q^2$ regime of the forthcoming BESIII measurement.  Finally, we emphasize that the $f_1(1420)$ form factors, particularly $F_{f_1\gamma^*\gamma^*}^{TL}(0,Q_2^2)$, can be obtained nearly model-independently from the polarized cross sections, as the $f_1(1420)$ channel dominates in the $\gamma\gamma^*\to K\bar{K}^*$ process around the $f_1(1420)$ resonance excitation and in the $Q_2^2$ range up to around $2$ GeV$^2$.

The presented model of the $\gamma\gamma^*\to K\bar{K}^*$ process in the $f_1(1420)$ energy region includes the necessary interference between the $s$-channel production of the $\eta(1475)$ and $f_1(1420)$ states. This interference is not accounted for in the Monte Carlo generator GaGaRes, typically used to simulate two-photon resonance production in $e^+e^-$ collisions. Therefore, our model can serve as a Monte Carlo generator for the BESIII measurement of the $\gamma\gamma^*\to K^+K^-\pi^0$ reaction, providing a tool to extract the $f_1(1420)$ TFFs.

\acknowledgments 
This work was supported by the Deutsche Forschungsgemeinschaft (DFG, German Research Foundation), in part through the Research Unit (Photonphoton interactions in the Standard Model and beyond, Projektnummer 458854507—FOR 5327), and in part through the Cluster of Excellence (Precision Physics, Fundamental Interactions, and Structure of Matter) (PRISMA$^+$ EXC 2118/1) within the German Excellence Strategy (Project ID 39083149).

\onecolumngrid
\appendix
\section{Cross section for the unpolarized single tagged $e^+ e^-\to e^+ e^- X$ process}\label{App:eeXsec}

In this appendix, we present the formulae for the cross section of the process $e(p_1) e(p_2) \to e(p_1')e(p_2') X$, where $X$ represents the produced hadronic state or system (e.g. $X=K\bar{K}^*$), in the c.m. system of the colliding beams. The general cross section formalism has been outlined in Ref.~\cite{Budnev:1975poe}.  Here, we specifically focus on the unpolarized single tagged process  $e^+ e^-\to e^+ e^- X$ and present an accessible formalism for the experimental analysis. 

The four-vector momenta of the incoming electron and positron in the $e^-e^+$ c.m. frame are denoted as $p_1(E,\bm{p}_1)$ and $p_2(E,-\bm{p}_1)$ with the beam energy $E=\sqrt{s}/2$ and $s=(p_1+p_2)^2$. The outgoing electron and positron have momenta $p_1'(E_1', \bm{p}_1')$ and $p_2'(E_2', \bm{p}_2')$, respectively. 
Regarding the unpolarized single tagged process, the lepton momentum $p_2'$ is detected, and the lepton momentum $p_1'$ goes undetected. This corresponds to the kinematical situation where the first photon is quasi real, and the second photon  has a finite virtuality. Then the photon four-momenta are 
\begin{equation}
	q_1 = p_1 - p_1',\quad q_2 = p_2 - p_2',
\end{equation} 
with the corresponding energies (in the $e^-e^+$ c.m. frame) and virtualities of the two photons expressed as 
\begin{equation}
    \omega_1 \equiv q_1^0 = E- E_1',\quad \omega_2 \equiv q_2^0 = E- E_2', 
\end{equation}
\begin{equation}
	Q_1^2 \equiv -q_1^2 \to 0, \quad 
	Q_2^2 \equiv -q_2^2 = 4E E_2'\sin^2\theta_2/2 + Q_{2,\text{min}}^2,
\end{equation}
where $\theta_2$ is the polar angle of the scattered lepton relative to the beam direction. The minimal value of the virtuality is given by 
\begin{equation}
	Q_{2,\text{min}}^2 \simeq m^2 \frac{\omega_2^2}{E\,E_2'},
\end{equation}
in the limit where $E_2'\gg m$ with $m$ being the electron mass. 
The $W^2=(q_1+q_2)^2$ stands for the squared invariant mass of the hadronic system. 

For the unpolarized single-tagged process, both the outgoing lepton four-momentum $p_2'$ and the invariant mass $W^2$ of the hadronic system are measured, which allows the energy $\omega_1$ of the quasi-real photon to be fixed as 
\begin{equation}
	\omega_1 = E\left( \frac{W^2+Q_2^2}{4E\omega_2 + Q_2^2}\right). 
\end{equation}
The differential unpolarized cross section of the single tagged measurement can be expressed as 
\begin{equation}\label{Eq:singletagXsec0}
\frac{d \sigma}{d \omega_2 \, d Q_2^2 \, d W^2}=\frac{1}{\left(\omega_2+\frac{Q_2^2}{4 E}\right) Q_2^2\left(W^2+Q_2^2\right)}
\biggl\{F_2^{++} \sigma_{T T}\left(W^2, Q_1^2=0, Q_2^2\right)+F_2^{00} \sigma_{TL}\left(W^2, Q_1^2=0, Q_2^2\right)\biggr\},
\end{equation}
where the virtual photon flux factors $F_2^{++}$ and $F_2^{00}$ are given by 
\begin{equation}
\begin{aligned}
F_2^{++} &  =\left(\frac{\alpha}{\pi}\right)^2 F_{1,\text{soft}} \left[1-\frac{1}{E}\left(\omega_2+\frac{Q_2^2}{4 E}\right)+\frac{1}{2E^2}\left(\omega_2+\frac{Q_2^2}{4 E}\right)^2\right],  \\
F_2^{00} & =\left(\frac{\alpha}{\pi}\right)^2  F_{1,\text{soft}}  \left[1-\frac{1}{E}\left(\omega_2+\frac{Q_2^2}{4 E}\right)\right] , 
\end{aligned}
\end{equation}
resulting from the integral over the quasi-real  photon virtuality $Q_1^2$, with $F_{1,\text{soft}}$ defined as 
\begin{equation}
F_{1,\text{soft}} = \left(1-\frac{\omega_1}{E}+\frac{\omega_1^2}{2 E^2}\right) \ln \frac{Q_{1, \text { max }}^2}{Q_{1, \text { min }}^2}-\left(1-\frac{\omega_1}{E}\right)\left(1-\frac{Q_{1, \text { min }}^2}{Q_{1, \text { max }}^2}\right).
\end{equation}
The bounds on the quasi-real photon virtuality $Q_1^2$ are  given by:  
\begin{equation}
	Q_{1,\text{min}}^2 \simeq m^2\frac{\omega_1^2}{E^2\left(1-\frac{\omega_1}{E}\right)},\quad 
	Q_{1,\text{max}}^2 \simeq 4E^2\left(1-\frac{\omega_1}{E}\right).
\end{equation}
Since the $\gamma \gamma^*\to X$ polarized cross sections, $\sigma_{TT}$ and $\sigma_{TL}$, in Eq.~\eqref{Eq:singletagXsec0} do not depend on $\omega_2$, one can perform the integration  over the experimentally accepted range of $\omega_2$ values, i.e. $\omega_{2, \text {min}}^{\text{exp.}}\leq \omega_2 \leq \omega_{2, \text {max}}^{\text{exp.}}$, which leads to the doubly differential cross section 
\begin{equation}
\frac{d \sigma}{d Q_2^2 \, d W^2}=\frac{\tilde{F}_2^{++}}{Q_2^2\left(W^2+Q_2^2\right)}\biggl\{\sigma_{TT}\left(W^2, Q_1^2=0, Q_2^2\right)+\varepsilon \,\sigma_{TL}\left(W^2, Q_1^2=0, Q_2^2\right)\biggr\},
\end{equation}
with the (dimensionless) integrated transverse virtual photon flux factor $\tilde{F}_2^{++}$  expressed as  
\begin{equation}
\tilde{F}_2^{++}=\int_{\omega_{2, \text{min}}^{exp.}}^{\omega_{2,\text{max}}^{\exp.}} \frac{d \omega_2}{\left(\omega_2+\frac{Q_2^2}{4 E}\right)} F_2^{++},	
\end{equation}
and the longitudinal photon polarization parameter $\varepsilon$ defined as 
\begin{equation}
	\varepsilon=\frac{1}{\tilde{F}_2^{++}} \int_{\omega_{2, \text{min}}^{\exp.}}^{\omega_{2, \text{max}}^{\exp.}} \frac{d \omega_2}{\left(\omega_2+\frac{Q_2^2}{4 E}\right)} F_2^{00}.
\end{equation}

\section{\label{App}Lorentz decomposition of the $\gamma\gamma^* \to VP$ reaction}

To perform the Lorentz decomposition of the non-resonant contribution of the $\gamma\gamma^*\to K\bar{K}^*(892)$ amplitude, Eq.~\eqref{Eq:nonresAmp}, we first derive the linearly-independent tensor basis for the fusion of one real-photon and one virtual-photon into a vector meson and a pseudoscalar meson: $\gamma(q_1)\gamma^*(q_2)\to V(p_1) P(p_2)$. Following the general recipe outlined by Bardeen, Tung, and Tarrach~(BTT)~\cite{Bardeen:1968ebo,Tarrach:1975tu}, and considering the on-shell condition of the final vector meson, we found $9$ independent tensors for the $\gamma\gamma^*\to V P$ reaction after applying the Schouten identity: 
\begin{align}
  T^1_{\mu\nu,\alpha} &= (q_1\cdot q_2)\epsilon_{\mu\nu\alpha \beta} (q_1+q_2)^\beta  + {(q_1)}_\nu \,\epsilon_{\alpha\mu\gamma\beta}(q_1)^\gamma {(q_2)}^\beta + {(q_2)}_\mu \,\epsilon_{\alpha\nu\gamma\beta}(q_1)^\gamma {(q_2)}^\beta ,\nonumber\\
  T^2_{\mu\nu,\alpha} &= (q_1\cdot q_2)\epsilon_{\mu\nu\alpha\beta}\Delta^\beta - {(q_1)}_\nu \epsilon_{\alpha\mu\gamma\beta} {(q_2)}^\gamma \Delta^\beta + {(q_2)}_\mu \epsilon_{\alpha \nu \gamma\beta} (q_1)^\gamma \Delta^\beta + g_{\mu\nu} \epsilon_{\alpha \sigma\gamma\beta} (q_1)^\sigma {(q_2)}^\gamma \Delta^\beta,\nonumber\\
 T^3_{\mu\nu,\alpha} &= (q_1-q_2)_\alpha \epsilon_{\mu\nu\gamma\beta} (q_1)^\gamma {(q_2)}^\beta, \nonumber\\
 T^4_{\mu\nu,\alpha} &= (q_1+q_2)_\alpha \epsilon_{\mu\nu\gamma\beta} (q_1)^\gamma {(q_2)}^\beta, \nonumber\\
 T^5_{\mu\nu,\alpha} &= (q_1\cdot q_2) {(q_1)}_\alpha \epsilon_{\mu\nu\gamma\beta}(q_1)^\gamma \Delta^\beta + (q_1\cdot q_2) {(q_2)}_\alpha \epsilon_{\mu\nu\gamma\beta}{(q_2)}^\gamma \Delta^\beta ,\nonumber\\
  &\quad + {(q_1)}_\alpha {(q_1)}_\nu \epsilon_{\mu\sigma\gamma\beta}(q_1)^\sigma {(q_2)}^\gamma \Delta^\beta + {(q_2)}_\alpha {(q_2)}_\mu \epsilon_{\nu \sigma\gamma\beta}(q_1)^\sigma {(q_2)}^\gamma \Delta^\beta, \\ 
 T^{6}_{\mu\nu,\alpha} &= \left( g_{\mu\nu} (q_1\cdot q_2) - {(q_2)}_\mu {(q_1)}_\nu \right) \epsilon_{\alpha \sigma\gamma\beta} (q_1)^\sigma {(q_2)}^\gamma \Delta^\beta,\nonumber\\
T^{7}_{\mu\nu,\alpha} &=  (q_2\cdot\Delta){(q_1)}_\nu \epsilon_{\alpha\mu\gamma\beta}(q_1)^\gamma\Delta^\beta + (q_1\cdot\Delta){(q_2)}_\mu \epsilon_{\alpha\nu\gamma\beta}{(q_2)}^\gamma\Delta^\beta \nonumber\\
  &\quad - (q_1\cdot q_2)\Delta_\mu \epsilon_{\alpha\nu\gamma\beta}{(q_2)}^\gamma \Delta^\beta - (q_1\cdot q_2)\Delta_\nu \epsilon_{\alpha\mu\gamma\beta}{(q_1)}^\gamma \Delta^\beta ,\nonumber\\
T^{8}_{\mu\nu,\alpha} &=  -(q_2\cdot\Delta){(q_1)}_\nu \epsilon_{\alpha\mu\gamma\beta}(q_1)^\gamma\Delta^\beta + (q_1\cdot\Delta){(q_2)}_\mu \epsilon_{\alpha\nu\gamma\beta}{(q_2)}^\gamma\Delta^\beta \nonumber\\
  &\quad - (q_1\cdot q_2)\Delta_\mu \epsilon_{\alpha\nu\gamma\beta}{(q_2)}^\gamma \Delta^\beta + (q_1\cdot q_2)\Delta_\nu \epsilon_{\alpha\mu\gamma\beta}{(q_1)}^\gamma \Delta^\beta ,\nonumber\\
T^{9}_{\mu\nu,\alpha} &=  -(q_2\cdot\Delta){(q_1)}_\nu \epsilon_{\alpha\mu\gamma\beta}(q_1)^\gamma {(q_2)}^\beta - (q_1\cdot\Delta){(q_2)}_\mu \epsilon_{\alpha\nu\gamma\beta}{(q_1)}^\gamma{(q_2)}^\beta \nonumber\\
  &\quad + (q_1\cdot q_2)\Delta_\mu \epsilon_{\alpha\nu\gamma\beta}{(q_1)}^\gamma {(q_2)}^\beta + (q_1\cdot q_2)\Delta_\nu \epsilon_{\alpha\mu\gamma\beta}{(q_1)}^\gamma {(q_2)}^\beta,\nonumber 
\end{align}
with $\Delta=p_1-p_2$. It is worth noting that the six independent tensors for the $\gamma\gamma\to V P$ reaction can be easily obtained by choosing $T^{3}$, $T^{4}$, $T^{6}$, $T^{7}$, $T^{8}$, and $T^{9}$ terms. Also, the current basis and the one obtained in Ref.~\cite{Ren:2022fhp} are identical after using a linear transformation.  

We applied the above tensor basis to perform the decomposition of the $\gamma\gamma^*\to K\bar{K}^*$ amplitude, Eq.~\eqref{Eq:nonresAmp}, and found the corresponding scalar functions:
\begin{align}
F_1(W^2,t,u)& = e^2\frac{g_{\gamma KK^*}}{m_K} \frac{1}{Q_2^2+W^2} \Biggl\{ F_K(-Q_2^2)\biggl[ \frac{2(m_K^2-M_{K^*}^2)+(Q_2^2+W^2)}{2(t-m_K^2)} +  \frac{2(m_K^2-M_{K^*}^2) - (Q_2^2+W^2)}{2(u-M_{K^*}^2)} \biggr] \nonumber\\
&\hspace{3.cm} + F_{\gamma^* KK^*}(-Q_2^2) \biggl[ -\frac{3(Q_2^2+W^2)}{2(t-M_{K^*}^2)} - \frac{Q_2^2+W^2}{2(u-m_K^2)}  \biggr]\Biggr\}, \nonumber\\
F_2(W^2,t,u) & = e^2\frac{g_{\gamma KK^*}}{m_K} \Biggl[ F_K(-Q_2^2) \frac{2}{u-M_{K^*}^2} - F_{\gamma^* KK^*}(-Q_2^2)\frac{2}{t-   M_{K^*}^2} \Biggr], \nonumber\\
F_3(W^2,t,u) & = e^2 \frac{g_{\gamma KK^*}}{m_K}\Biggl\{ F_K(-Q_2^2)\frac{1}{Q_2^2+W^2}\biggl[ \frac{m_K^2-M_{K^*}^2}{2(t-m_K^2)} + \frac{m_K^2-M_{K^*}^2}{2(u-M_{K^*}^2)}  \biggr] \nonumber\\
&\hspace{2cm} + F_{\gamma^* KK^*}(-Q_2^2)  \frac{1}{Q_2^2+W^2} \biggl[ \frac{m_K^2-M_{K^*}^2}{2(u-m_K^2)} + \frac{m_K^2-M_{K^*}^2 - 4 (Q_2^2+W^2)}{2(t-M_{K^*}^2)}\biggr]\Biggr\}, \nonumber \\
F_4(W^2,t,u) & = e^2 \frac{g_{\gamma KK^*}}{m_K} \Biggl\{ F_K(-Q_2^2) \frac{1}{Q_2^2+W^2} \biggl[ \frac{m_K^2-M_{K^*}^2- (Q_2^2+W^2)}{2(t-m_K^2)}  + \frac{m_K^2-M_{K^*}^2- 5(Q_2^2+W^2)}{2(u-M_{K^*}^2)} \biggr] \nonumber\\
&\hspace{2cm} + F_{\gamma^* KK^*}(-Q_2^2) \frac{1}{Q_2^2+W^2}\biggl[ \frac{m_K^2-M_{K^*}^2 - (Q_2^2+W^2)}{2(u-m_K^2)} + \frac{m_K^2-M_{K^*}^2 - 5(Q_2^2+W^2)}{2(t-M_{K^*}^2)} \biggr] \Biggr\}, \nonumber\\
F_5(W^2,t,u) & = 0, \\
F_6(W^2,t,u) & = 0, \nonumber \\
F_7(W^2,t,u) & = e^2\frac{g_{\gamma KK^*}}{m_K} \frac{1}{(Q_2^2+W^2)} \Biggl\{ F_K(-Q_2^2) \biggl[ \frac{1}{t-m_{K}^2} + \frac{1}{u-M_{K^*}^2} \biggr] + F_{\gamma^* KK^*}(-Q_2^2) \biggl[ \frac{1}{t-M_{K^*}^2} +\frac{1}{u-m_K^2}\biggr] \Biggr\}, \nonumber \\
F_8(W^2,t,u) & = - e^2\frac{g_{\gamma KK^*}}{m_K} \frac{1}{(Q_2^2+W^2)} \Biggl\{ F_K(-Q_2^2) \biggl[ \frac{1}{t-m_{K}^2} + \frac{1}{u-M_{K^*}^2} \biggr]  - F_{\gamma^* KK^*}(-Q_2^2) \biggl[ \frac{1}{t-M_{K^*}^2} +\frac{1}{u-m_K^2}\biggr] \Biggr\}, \nonumber \\
F_9(W^2,t,u) & = -e^2 \frac{g_{\gamma KK^*}}{m_K} \frac{1}{ (Q_2^2+W^2)} \Biggl\{  F_K(-Q_2^2) \biggl[ \frac{1}{t-m_{K}^2}+ \frac{1}{u-M_{K^*}^2} \biggr] - F_{\gamma^* KK^*}(-Q_2^2) \biggl[ \frac{1}{t-M_{K^*}^2} +\frac{1}{u-m_K^2}\biggr] \Biggr\}. \nonumber
\end{align}

\twocolumngrid

\bibliographystyle{apsrev4-2}
\bibliography{ref_gagatoKKbars.bib}

\end{document}